\documentclass[aps,pra,twocolumn,superscriptaddress,preprintnumbers,amsmath,amssymb]{revtex4-2}
\setcitestyle{super,sort&compress}
\usepackage{enumitem}
\setlist[itemize]{itemsep=0.15em, topsep=0.15em, parsep=0em, partopsep=0em}
\usepackage{yhmath}
\usepackage{makecell}
\usepackage{braket}
\usepackage{mathtools}
\usepackage{amsmath,amsthm,titlesec}
\usepackage{physics}
\usepackage[T1]{fontenc}
\usepackage[markup=underlined]{changes}
\usepackage[normalem]{ulem}
\usepackage{upgreek}
\usepackage{threeparttable} 
\usepackage{multirow}
\usepackage[modulo]{lineno}
\usepackage{siunitx}
\usepackage{comment}
\newtheorem{Theorem}{Theorem}

\newtheorem{Lemma}{Lemma}
\newcommand{\beq}{\begin{equation}}
\newcommand{\eeq}{\end{equation}}
\newcommand{\beqa}{\begin{eqnarray}}
\newcommand{\eeqa}{\end{eqnarray}}
\usepackage{float}
\usepackage{subfigure}
\usepackage{multirow}
\usepackage[colorlinks=true, linkcolor=blue, citecolor=blue, urlcolor=blue]{hyperref} 
\makeatletter
\g@addto@macro\normalsize{%
\setlength\abovedisplayskip{4pt plus 1pt minus 1pt}
\setlength\belowdisplayskip{4pt plus 1pt minus 1pt}
\setlength\abovedisplayshortskip{2pt plus 1pt minus 1pt}
\setlength\belowdisplayshortskip{2pt plus 1pt minus 1pt}
\setlength{\jot}{2pt plus 1pt minus 1pt}
}%
\makeatother

\begin{document}

\title{The Riemann Hypothesis Emerges in Dynamical Quantum Phase Transitions}
\author{ShiJie Wei}
\email{weisj@baqis.ac.cn}
\altaffiliation{These authors contributed equally to this work.}
\affiliation{Beijing Academy of Quantum Information Sciences, Beijing 100193, China}

\author{Yue Zhai}

\altaffiliation{These authors contributed equally to this work.}
\affiliation{Shenzhen Institute for Quantum Science and Engineering, Southern University of Science and Technology, Shenzhen 518055, China}
\affiliation{International Quantum Academy, Futian District, Shenzhen, Guangdong 518048, China}

\author{Quanfeng Lu}
\altaffiliation{These authors contributed equally to this work.}
\affiliation{State Key Laboratory of Low-Dimensional Quantum Physics and Department of Physics, Tsinghua University, Beijing 100084, China}

\author{Wentao Yang}
\affiliation{State Key Laboratory of Low-Dimensional Quantum Physics and Department of Physics, Tsinghua University, Beijing 100084, China}
\author{Pan Gao}
\affiliation{Beijing Academy of Quantum Information Sciences, Beijing 100193, China}
\author{Chao Wei}
\affiliation{International Quantum Academy, Futian District, Shenzhen, Guangdong 518048, China}
\affiliation{Shenzhen Institute for Quantum Science and Engineering, Southern University of Science and Technology, Shenzhen 518055, China}

\author{Junda Song}
\affiliation{International Quantum Academy, Futian District, Shenzhen, Guangdong 518048, China}
\affiliation{Shenzhen Institute for Quantum Science and Engineering, Southern University of Science and Technology, Shenzhen 518055, China}

\author{Franco Nori}

\affiliation{Center for Quantum Computing, RIKEN, Wakoshi, Saitama, 351-0198, Japan}
\affiliation{Quantum Research Institute, The University of Michigan, Ann Arbor, 48109-1040, MI, USA}

\author{Tao Xin}\email{xintao@iqasz.cn}
\affiliation{International Quantum Academy, Futian District, Shenzhen, Guangdong 518048, China}

\author{Guilu Long}\email{gllong@tsinghua.edu.cn}
\affiliation{State Key Laboratory of Low-Dimensional Quantum Physics and Department of Physics, Tsinghua University, Beijing 100084, China}
\affiliation{Tsinghua National Laboratory for Information Science and Technology, Beijing 100084, People’s Republic of China}
\affiliation{Beijing Academy of Quantum Information Sciences, Beijing 100193, China}

\begin{abstract}

The Riemann Hypothesis (RH), one of the most profound unsolved problems in mathematics, concerns the nontrivial zeros of the Riemann zeta function. Establishing connections between the RH and  physical phenomena could offer new perspectives on its physical origin and verification. Here, we establish a direct correspondence between the nontrivial zeros of the zeta function and dynamical quantum phase transitions (DQPTs) in two realizable  quantum systems, characterized by the averaged accumulated phase factor and the Loschmidt amplitude, respectively. This  precise correspondence reveals that the RH can be viewed as the emergence of DQPTs at a specific  temperature. We experimentally demonstrate this correspondence on a five-qubit spin-based system and further propose an universal quantum simulation framework for efficiently realizing both systems with polynomial resources, offering a quantum advantage for numerical verification of the RH. These findings uncover an intrinsic link between nonequilibrium critical dynamics and the RH, positioning quantum computing as a powerful platform for exploring one of mathematics’ most enduring conjectures and beyond.

\end{abstract}

\maketitle
\section{introduction}

\begin{figure*}[hbt!]
    \begin{center}
    \includegraphics[width=2\columnwidth]{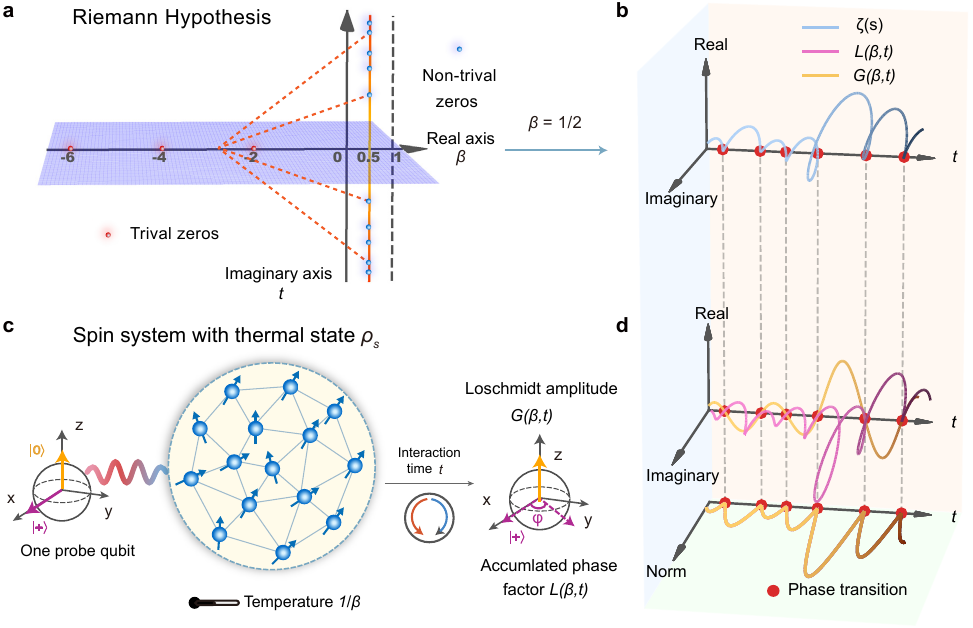}
    \end{center}
    \vspace{-0.50cm}
    \caption{\textbf{
    Correspondence between the Riemann zeta function and physical many-body systems.} 
    \textbf{a}
    The RH in the complex plane $(\beta, t)$. It states that all nontrivial zeros of the  zeta function lie on the critical line 
    with real part $\beta=1/2$. The distribution of the zeros of the zeta function is symmetric with respect to reflection across the real axis. 
    \textbf{b} The  zeta function $\zeta(1/2+i t)$ along the critical line. 
    Each zero crossing, marked by a  red dot,  corresponds to  a nontrivial Riemann zero.
    \textbf{c} 
    Construction of two quantum systems, each consisting of a single probe spin coupled to a many-body system initialized in a thermal state at inverse temperature $\beta$. The composite system subsequently evolves under the system interactions with time-reversal symmetry. The blue arrow  and red arrow represent the forward time evolution and backward evolution respectively.  
   \textbf{d} 
    The dynamics of the accumulated phase factor ${\cal L}(\beta,t)$ and the Loschmidt amplitude  ${\cal G}(\beta,t)$ reproduce key features of the zeta function, exhibiting periodic vanishing-and-revival oscillations characteristics of DQPTs. The norms of both observables show correspondence with the norm of the  zeta function.
    } 
    \vspace{-0.50cm}
    \label{Fig1}
\end{figure*}
The Riemann zeta function is central to analytic number theory and has profound implications in mathematical physics. It admits a Dirichlet series representation valid for $ \Re(s) > 0, \Re(s) \neq 1$~\cite{milgram2013integral}:
\begin{equation} \label{eq:zeta_function}
\zeta(s) = \frac{1}{1 - 2^{1-s}} \sum_{n=1}^\infty \frac{(-1)^{n+1}}{n^s}.
\end{equation}
 
The nontrivial zeros of the zeta function,  conjectured by the celebrated Riemann Hypothesis (RH) to all lie on the critical line $\Re(s)=\tfrac{1}{2}$, remain one of the most profound unsolved problems in mathematics. The RH underpins more than a thousand theorems in mathematics and forms the foundation of the modern RSA cryptographic system, for example, their distribution encodes the fine structure of prime numbers~\cite{bui2011large}.
Despite extensive numerical evidence~\cite{odlyzko200110,Bober03042018}, the RH remains unproven. Any new perspective on the  zeta function is therefore of great significance, regardless of whether it ultimately leads to a proof or refutation of the RH.

  Beyond its mathematical significance, the Riemann zeta function has long inspired deep physical analogies~\cite{schumayer2011colloquium}, sparking the search for a possible physical origin of the RH. The most influential idea in this direction is the Hilbert–Pólya conjecture~\cite{berry2005riemann,berry1992new},  which suggests that the nontrivial zeros of the zeta function might correspond to the eigenvalues of an yet-unknown quantum Hamiltonian, motivating searches for such an operator~\cite{srednicki2011nonclassical,bender2017hamiltonian} and for signatures in diverse physical contexts, including relativistic amplitudes~\cite{remmen2021amplitudes}, chaotic quantum scattering~\cite{gutzwiller1983stochastic,bhaduri1995phase}, quantum field theory~\cite{spector1990supersymmetry}, and random matrix theory~\cite{hughes2000random}. Significant progress~\cite{mack2010riemann,feiler2013entanglement,feiler2015dirichlet} has linked quantum systems to the Riemann zeta function through tailored state preparation, engineered spectra, and observable correspondences with its zeros. The search for a genuine physical realization, deeper connections and possible origins of the RH  remains unknown.

Here, we propose a physical origin for the RH by demonstrating that the non-trivial zeros of the Riemann zeta function correspond directly to dynamical quantum phase transitions (DQPTs) in two distinct constructed quantum systems~\cite{heyl2018dynamical,heyl2013dynamical,karrasch2013dynamical,heyl2017dynamical,bhattacharya2017mixed,lang2018dynamical,muniz2020exploring,zhou2018dynamical}. 
 Their physical realization starts with both models initialized in thermal equilibrium. We then drive them out of equilibrium by quenching with specific interaction Hamiltonians, which are engineered to imprint the zeta function's properties onto measurable observables. The average accumulated phase factor in the first system and the Loschmidt amplitude in the second both vanish at evolution times corresponding to the nontrivial zeros, revealing DQPTs. 
 Furthermore, we introduce a quantum computational framework to efficiently simulate these dynamics on digital quantum computers, offering a clear quantum advantage for probing both DQPTs and the RH. This approach not only enables the experimental observation of DQPTs tied to the zeros but also allows for the numerical exclusion of such critical behavior at non-zero points, thereby providing a physical pathway to investigate the RH.

As a proof of principle, we 
implement the first system experimentally on a five-qubit nuclear spin quantum processor. The measured coherence dynamics 
reproduce the oscillatory behavior of the zeta function and exhibit clear signatures of DQPTs aligned with its nontrivial zeros. For the second system, suitable for exploring higher-order zeros, large-scale numerical simulations confirm the correspondence up to the $10^{12}$-th zero. 
Moreover, our gate-based framework constructs both models with polynomial resources and achieves at least a quadratic speedup over classical verification of RH zeros. These results establish a concrete bridge between number theory, quantum physics, and quantum technology, opening scalable  pathways toward one of mathematics' most enduring mysteries and hinting at a deeper quantum structure underlying arithmetic.

\section{Results}

As shown in Fig. \ref{Fig1}c, we study two distinct quantum dynamical processes, each initialized in a thermal state.

In both cases,
a single probe qubit is coupled to the system, and the combined system evolves under an interaction Hamiltonian for a duration $t$.  The average accumulated phase factor ${\cal L}(\beta,t)$ in the first setup directly encodes the behavior of $\zeta(s)$ in the thermodynamic limit, while the Loschmidt amplitude ${\cal G}(\beta,t)$ in the second  captures the same feature in the joint thermodynamic and long-time limits. Remarkably, at $\beta = \frac{1}{2}$, both systems exhibit  DQPTs, characterized by periodic vanishing-and-revival oscillations in the observables, which maps one-to-one to the nontrivial zeros. If the RH holds, such periodic revivals occur exclusively at this critical line $\beta = \frac{1}{2}$ in the first system. 
This  correspondence indicates that RH can be explained as the  DQPTs  emerges only at inverse temperature  $\beta = \frac{1}{2}$ in a  specific quantum many body system. Moreover, the quantum systems exhibit time-reversal symmetry, corresponding to the reflection symmetry of the zeta function. As summarized in Table~\ref{correspondence}, we establish correspondence between  the engineered quantum system and mathematical features of the zeta function. 
Together, these results establish a direct correspondence between the analytic features of the zeta function (Fig. \ref{Fig1}b) and accessible quantum dynamics (Fig. \ref{Fig1}d). 
We describe the construction of these quantum dynamics in the following sections.

\begin{table}[!ht]
\caption{\textbf{Correspondence between quantum systems and the Riemann zeta function.} Engineered quantum dynamics enable direct mapping of the zeta function, its nontrivial zeros, and its  reflection symmetry.}
\label{tab:scaling}
\centering
\renewcommand{\arraystretch}{2.5}
\begin{tabular}{|c|c|}
\Xhline{0.7pt}
\makecell{\textbf{Physical observables} \\ ${\cal L}( \beta,t)$ and ${\cal G}(\beta,t)$} & \makecell{\textbf{Zeta function} \\ $\zeta(s)$ }\\
\Xhline{0.7pt}
\makecell{Accumulated phase factor ${\cal L}( \beta, t)$ \\ vanishes at $\beta = 1/2 $ as $t$ increases} & Zeros of $\zeta(s)$ \\
\hline
\makecell{Loschmidt amplitude ${\cal G}(\beta,t)$ \\ vanishes at  $\beta = 1/2 $  as $t$ increases~} & Zeros of $\zeta(s)$ \\
\hline
DQPT occurs exclusively at  $\beta = 1/2 $ & RH\\
\hline
Time-reversal invariance & \makecell{Reflection \\ ~symmetry of $\zeta(s)$ ~} \\
\Xhline{0.7pt}
\end{tabular}
\label{correspondence}
\end{table}

\subsection{Probe Spin Coherence Signaling RH}\label{model1}
We consider a quantum system with Hamiltonian $\mathcal{H}_0 = \sum_{n=1}^{N} E_n \ket{n}\bra{n}$  and a logarithmic energy
spectrum  $E_n = \log n$~\cite{mack2010riemann}. 
The system is initialized in the thermal equilibrium state 
$\rho_s = \tfrac{\mathrm{e}^{-\beta \mathcal{H}_0}}{\mathcal{Z}(\beta,\mathcal{H}_0)}$, with partition function 
$\mathcal{Z}(\beta,\mathcal{H}_0) = \sum_{n=1}^{N} n^{-\beta}$. 
Constructing such a logarithmic energy spectrum is central to realizing the quantum system. Previous studies~\cite{gleisberg2013factorization,cassettari2023holographic} have typically employed analog simulation approaches, designing specific potential landscapes to approximate a logarithmic energy distribution. In our experiments, we prepare the corresponding thermal state directly by engineering level populations of density matrices in the nuclear spin quantum processor. More importantly, we present a universal quantum algorithm that prepares states and Hamiltonians with a logarithmic spectrum to arbitrary precision using only polynomial resources (see Sec.~\ref{QC}). 

We define the average accumulated phase factor  as ${\cal L}(\beta,t)=\operatorname{Tr}\left[\rho_s \mathrm{e}^{-i \mathcal{H}_0t} \sigma^1_z\right]$, where $\mathrm{e}^{-i \mathcal{H}_0t} \sigma^1_z$ imprints a phase factor $(-1)^n \mathrm{e}^{-iE_nt}$ on the $n$-th population of $\rho_s$. 
This yields
\begin{align}\label{eq:L}
    {\cal L}(\beta,t)=&\sum_{n=1}^{N}(-1)^n\mathrm{e}^{-iE_nt} \mathrm{e}^{-\beta E_n}/{\cal Z}(\beta,\mathcal{H}_0)\notag\\
    =&-\sum_{n=1}^{N}(-1)^{n+1}n^{-\beta-it}/{\cal Z}(\beta,\mathcal{H}_0).
\end{align}

In the thermodynamic limit $N \to \infty$, 
$\mathcal{L}(\beta,t)$ is directly related to the zeta function $\zeta(\beta + i t)$ as
\begin{eqnarray}\label{eq:relation1}
    {\cal Z}(\beta,\mathcal{H}_0){\cal L}(\beta,t)\xrightarrow{N\rightarrow\infty}(2^{(1-(\beta+it))}-1)\zeta(\beta+it).
\end{eqnarray}
  Hence, $\mathcal{L}(\beta,t)$ vanishes precisely when $\zeta(\beta + i t)$ has a nontrivial zero. These zeros signal DQPTs in mixed states marked by non-analyticities in time \cite{heyl2017dynamical}. 
  We next establish a correspondence between these dynamical transitions and their equilibrium analogues.
 

Through the Choi–Jamiołkowski isomorphism, $\rho_s$ can be mapped as
$|\rho_s\rangle=\sum_{n=1}^{N} n^{-\beta/2}|n\rangle |n\rangle_a $, where $|n\rangle_a$ denotes an orthonormal basis in an ancillary Hilbert space  isomorphic to the physical one.
Under time evolution, the state evolves as $|\rho(t)\rangle = (\mathrm{e}^{-i \mathcal{H}_0t} \otimes \mathbb{I}_a) |\rho_s\rangle.$
and the accumulated phase factor becomes
$\mathcal{L}_p(\beta,t)=\langle\rho_s | \mathrm{e}^{-i \mathcal{H}_0t} \sigma^1_z|\rho_s\rangle$.
Consider an equilibrium system with boundary conditions imposed at two ends separated by a distance $R$,
the boundary partition function in equilibrium statistical mechanics~\cite{heyl2018dynamical, leclair1995boundary} is given by
\begin{equation}
   \mathcal{Z}_{B}(\beta,R,\mathcal{H}) =\langle \psi_a | \mathrm{e}^{-R \mathcal{H}} | \psi_b \rangle,
\end{equation}
where $|\psi_a\rangle=\tfrac{1}{C}\textstyle\sum\limits_{n=1}^{N} n^{-\tfrac{\beta}{2}}|n\rangle|n\rangle_a$ and $|\psi_b\rangle= \sigma^1_z|\psi_a\rangle$ encode the boundary states and $\mathcal{H}$ denotes the bulk Hamiltonian. 
By analytically continuing to complex time plane $R = it$, we obtain

\begin{equation}\label{eq:ZL}
    \mathcal{Z}_{B} ( \beta,it,\mathcal{H}_0\otimes \mathbb{I}_a)= \langle \psi_a | \mathrm{e}^{-i \mathcal{H}_0 t}\otimes \mathbb{I}_a | \psi_b \rangle=\mathcal{L}(\beta,t).
\end{equation}
Thus, $\mathcal{L}(\beta,t)$ serves as a boundary partition function analytically continued to complex time. 
In the thermodynamic limit, the free-energy density is defined as~\cite{heyl2013dynamical,peng2015experimental,francis2021many}
\begin{align}\label{eq:free} \mathcal{F}_1(\beta,t) &= - \lim_{N \to \infty} \frac{1}{\log N} \ln |\mathcal{Z}_{B}( \beta,it,\mathcal{H}_0\otimes \mathbb{I}_a)|\notag \\
&=- \lim_{N \to \infty} \frac{1}{\log N} \ln |\mathcal{L}( \beta,t)|.
\end{align}
Singularities in $\mathcal{F}_1(\beta,t)$ occurs precisely at the nontrivial zeros, signaling the occurrence of DQPTs~(see Methods~\ref{Methods:free energy} for a detailed derivation).
If the RH holds, such singularities occur exclusively along the critical line $\beta =1/2$, 
analogous to Lee–Yang zeros in thermal systems~\cite{wei2012lee, peng2015experimental,francis2021many}.
For other temperatures $\beta \neq 1/2$, no such phenomenon arises.

According to equation \eqref{eq:ZL}, the same correspondence 
can be realized in a pure-state setting and  we omit the detailed derivation here for brevity.

To render $\mathcal{L}(\beta,t)$ experimentally accessible,
we introduce a probe spin initialized in the superposition state $\ket{+} = (\ket{\uparrow} + \ket{\downarrow})/\sqrt{2}$. The probe is coupled to the system via
$\mathcal{H}_{c1} = \lambda \ket{\downarrow}\bra{\downarrow} \otimes \sigma^1_z$, where $\lambda$ denotes the interaction strength. For an interaction time $t_0$ satisfying $\lambda t_0 = \pi$, this operation 
imprints a phase $(-1)^n$ on the $n$-th component of $\rho_s$. The joint system then evolves under the interaction Hamiltonian $\mathcal{H}_{c2} = \ket{\downarrow}\bra{\downarrow} \otimes \mathcal{H}_0$ for a duration time $t$. The expectation value $\langle \sigma_x \rangle + i \langle \sigma_y \rangle$ of the probe spin directly yields $\mathcal{L}(\beta,t)$, providing experimental access to the dynamical signatures.

\begin{figure*}[hbt!]
    \begin{center}
    \includegraphics[width=2.1\columnwidth]{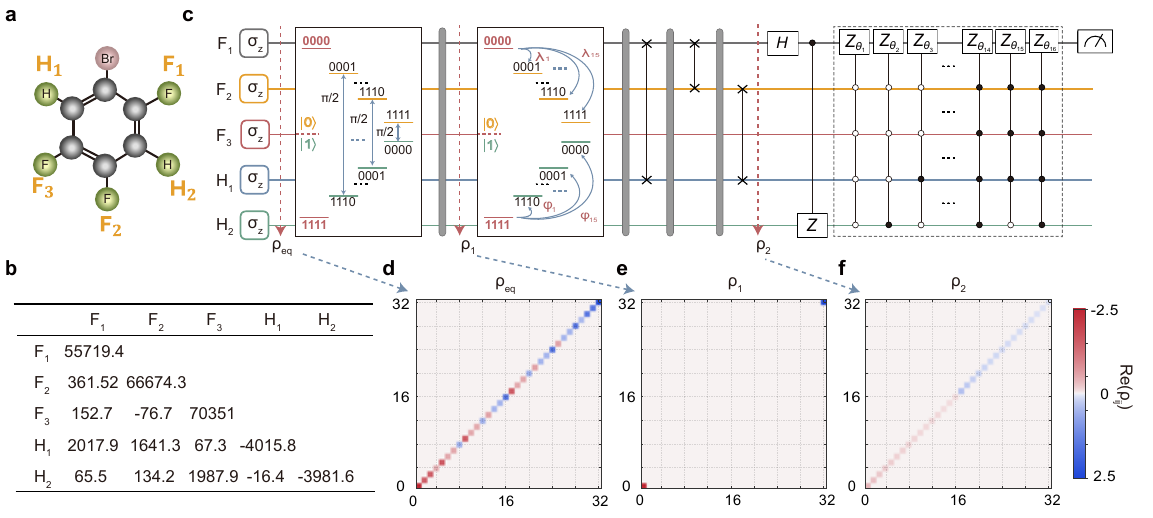}
    \end{center}
    \vspace{-0.50cm}
    \caption{\textbf{Physical system and quantum circuit for simulating the Riemann zeta function.} \textbf{a,} Molecular structure of the 1-bromo-2,4,5-trifluorobenzene. The five nuclear spins are encoded as qubits, with the F$_1$ spin serving as the probe qubit to extract the average accumulated phase factor.
    \textbf{b,} Measured Hamiltonian parameters. The chemical shifts $\nu_i$ are shown on the diagonal, while the off-diagonal elements represent the coupling strengths $(J_{ik}+2D_{jk})$, both in units of Hz. \textbf{c,} Quantum circuit implementation, including thermal state preparation, controlled dynamical evolution, and coherence measurement. $H$ denotes the Hadamard gate acting on the probe qubit. Coherent operations for state preparation and multi-qubit quantum gates are realized using shaped control pulses. The parameters $\lambda_i$ and $\psi_i$ for population transfer are determined by $\beta$, and the parameters $\theta_i$ are determined by $t$. \textbf{d, e, f,} Effective population evolution during state preparation. Here, we employ the traceless deviation density matrix to describe the state evolution.} 
    \vspace{-0.50cm}
    \label{Fig2}
\end{figure*}

\subsection{Loschmidt echo corresponding to the RH}\label{model2}

We further construct a concrete quantum system  
whose Loschmidt amplitude is directly linked to the zeta function, enabling the verification its large  zeros. 
The system is initialized in the thermal equilibrium state $\rho_s$ of the Hamiltonian $\mathcal{H}_0$, and is coupled to a probe spin, 
forming the composite mixed state $\rho_0 = \ket{\uparrow}\bra{\uparrow} \otimes \rho_s$. 
The joint system then evolves  under the  time-dependent Hamiltonian 
$\mathcal{H}_{\rm c} = \sigma_x \otimes (\mathcal{H}_0-\dot{\theta}(t')\mathcal{I})$ for $0\leq t'\leq t$, with time-evolution operator
$U(t') = \exp(-i\textstyle\int_{0}^{t}\mathcal{H}_{\rm c} dt')$. 
Here $\theta(t')= \Im(\log(\tfrac{1}{4} + \tfrac{i t'}{2})) - \tfrac{t'}{2}\log \pi$ is the Riemann-Siegel theta function~\cite{de2011high}, and its time derivative is $\dot{\theta}(t')=\tfrac{1}{2}\Re(\psi(\tfrac14+\tfrac{i t'}{2}))-\tfrac{1}{2}\log\pi$, with $\psi$ denoting the digamma function.\par
The generalized Loschmidt amplitude (GLA) is defined as the overlap of the initial  density matrices with the time-evolution operator~\cite{heyl2017dynamical}. This yields

\begin{align}\label{Loschmidt amplitude}
\mathcal{G}(\beta,t)\!\!\!\;
=&\mathrm{Tr}[\mathrm{e}^{-i\textstyle\int_{0}^{t}\mathcal{H}_{\rm c} dt'}\rho_0]\notag\\
=&\tfrac{1}{2\mathcal{Z}(\beta,H_0)}\;\!\!(\mathrm{e}^{i \theta(t)}\!\textstyle\sum\limits_{\mathclap{n=1}}^{\mathclap{N}}n^{-\beta - it}\!\;\!\!\;\!\! +\!\;\!\!\mathrm{e}^{-i \theta(t)}\textstyle\!\sum\limits_{\mathclap{n=1}}^{\mathclap{N}}n^{-\beta + it}).
\end{align} 
For $N=\sqrt{t/{2\pi}}$, ${\cal G}(\tfrac{1}{2},t)$ matches the Hardy 
Z-function $Z(t)=\mathrm{e}^{i\theta t}\zeta(\tfrac{1}{2}+it)$, which admits a
Riemann-Siegel approximation~\cite{de2011high},
\begin{equation}
   Z(t) =\mathrm{e}^{i\theta(t)}\textstyle\sum\limits_{\mathclap{n=1}}^{\mathclap{N}} n^{-\frac{1}{2}-it} 
         +\mathrm{e}^{-i\theta(t)}\textstyle\sum\limits_{\mathclap{n=1}}^{\mathclap{N}} n^{-\frac{1}{2}+it} 
         + \mathcal{O}(t^{-\frac{1}{4}}).
   \label{eq:RS_formula}
\end{equation}
In the joint thermodynamic and long-time limit,
\begin{equation}\label{eq:C}
{\cal G}(\tfrac{1}{2},t) \xrightarrow[N=\sqrt{t/2\pi}]{t\to\infty} \frac{\mathrm{e}^{i\theta(t)}\zeta(\tfrac{1}{2} + it)}{2{\cal 
Z}(\tfrac{1}{2},\mathcal{H}_0)} .
\end{equation}

The modulus of the Loschmidt amplitude, ${\cal B}(t) = |{\cal G}(t)|^2$, defines the Loschmidt echo (LE), from which the Loschmidt rate (or dynamical free energy) is obtained as 
$\mathcal{F}_2(t) = -\textstyle\lim\limits_{ N \to \infty} \frac{1}{\log N} \ln {\cal B}(t)$. 
Nonanalyticities in $\mathcal{F}_2(t)$ at critical times $t_c$, where ${\cal G}(\beta,t_c)=0$, signal DQPTs~\cite{heyl2013dynamical,heyl2017dynamical}. 
At $\beta=\tfrac{1}{2}$,  the LE exhibits a 
vanishing-and-revival dynamic, vanishing at the critical times $t_c$ that correspond to the nontrivial Riemann zeros. 
As the evolution time $t$ increases, this correspondence becomes increasingly accurate. 

We demonstrate that the GLA can be accessed via the expectation value of $\sigma_z$ on the probe spin (see Methods~\ref{Methods:Measure LA}).  Extended Data Fig. 1a shows numerical simulations of a three-spin system  reproducing the zeros of the GLA corresponding to nontrivial Riemann zeros with imaginary parts in the range $[420,450]$. Extending to a ten-spin system enables access to zeros with imaginary parts in $[6.595 \times 10^6,6.595 \times 10^6 + 10]$, corresponding to the $\num{13502344}$-th to $\num{13502366}$-th zeros, as shown in  Extended Data Fig. 1b. An eighteen-spin system further captures the DQPTs linked to the $10^{12}$-th and nearby zeros as shown in  Extended Data Fig. 1c. Notably, the deviation between the estimated zeros and exact zeros decreases with increasing $t$, indicating improved precision of correspondence at larger Riemann zeros. In addition, we consider a pure initial state $\frac{1}{C}\sum_{n=1}^{N} n^{-\beta/2}|0\rangle|n\rangle$ evolving under the system Hamiltonian $\mathcal{H}_{\rm c}$. This system likewise exhibits DQPTs, characterized by nonanalytic behavior in the  Loschmidt amplitude, whose zeros correspond to the nontrivial zeros of the zeta function.


\begin{figure*}[hbt!]
    \begin{center}
    \includegraphics[width=2.1\columnwidth]{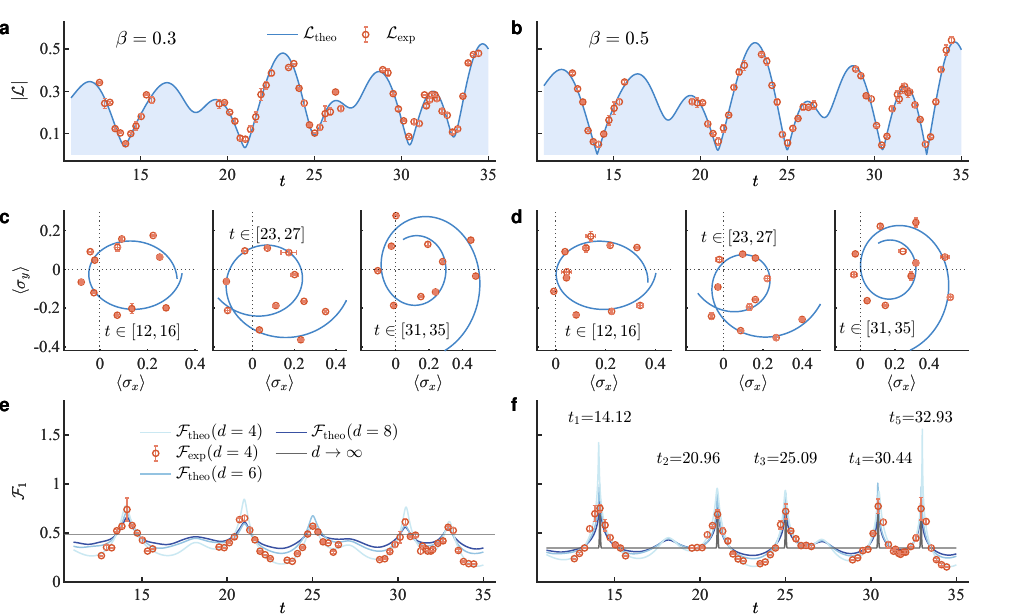}
    \end{center}
    \vspace{-0.50cm}
    \caption{\textbf{The coherence dynamics ${\cal L}(\beta, t)$ of the probe qubit and the non-trival Riemann zeros. }  \textbf{a,} Measured probe spin coherence ${\cal L}(0.3, t)$ versus evolution time $t$. \textbf{b,} ${\cal L}(0.5, t)$ versus time $t$ for $\beta = 0.5$. Solid lines represent numerically calculated values of the probe spin coherence. 
    \textbf{c, d,} The first, third, and fifth Riemann zeros identified from the zeros of the probe spin coherence, where both the real part $\langle \sigma_x \rangle$ and the imaginary part $\langle \sigma_y \rangle$ simultaneously vanish. 
    \textbf{e, f,} The corresponding free-energy density computed using Eq. (\ref{eq:free}). $d=\log N$ reflects the system’s effective number of degrees of freedom.
    Theoretically predicted Riemann zeros are indicated by the values on th peaks for comparison. For $\beta = 0.5$, which corresponds to the critical line of the RH, the dynamics of the probe spin coherence exhibit multiple zeros, each matching one-to-one with Riemann zeros. In contrast, for $\beta = 0.3$, which lies outside the RH, no zeros are observed in the spin coherence dynamics. Here and elsewhere, error bars represent 1$\sigma$ confidence intervals coming from the repeated experiments. These findings establish our approach as a powerful tool for investigating the RH and its zeros within a controlled quantum many-body system. 
    } 
    \label{Fig3}
\end{figure*}

\subsection{Experimental verification using nuclear spins}

The connection between physical quantum systems and the RH offers a novel method for experimentally probing the behavior of the zeta function through controlled  Hamiltonian dynamics. As a proof-of-principle demonstration, we realized a five-qubit system on a nuclear magnetic resonance (NMR) platform. The qubits are encoded in the nuclear spins of a 1-bromo-2,4,5-trifluorobenzene molecule, partially aligned in the nematic liquid crystal solvent N-(4-methoxybenzaldehyde)-4-butylaniline (MBBA)~\cite{exp1,exp2,exp3}. The molecule consists of three ${}^{19}\mathrm{F}$ nuclei and two ${}^{1}\mathrm{H}$ nuclei. A single nuclear spin serves as the probe qubit, with the remaining spins constituting a sixteen-dimensional working system. The effective Hamiltonian of this five-qubit system in the rotating frame is given by
\begin{equation}
\mathcal{H}_{\mathrm{NMR}} = \sum_{j=0}^{4} \pi \nu_j \frac{\sigma_z^{j}}{2}+\sum_{j<k} \frac{\pi}{2} \left( J_{jk} + 2D_{jk} \right) \sigma_z^{j} \sigma_z^{k},
\label{eq:nmr_hamiltonian}
\end{equation}
where $\sigma_z^{j}$ is the Pauli-$z$ operator on the $j$-th spin, $\nu_j$ denotes the chemical shift of the $j$-th spin, and $(J_{jk} + 2D_{jk})$ represents the effective coupling strength between spins $j$ and $k$. Here, $J_{jk}$ and $D_{jk}$ correspond to the scalar and residual dipolar couplings, respectively. The molecular structure and Hamiltonian parameters are illustrated in Figs.~\ref{Fig2}a and ~\ref{Fig2}b, respectively. All experiments were implemented on a Bruker Avance III 600 MHz spectrometer at $T = 305$ K.

There are two major challenges in experimentally realizing this constructed Hamiltonian dynamics. The first is the preparation of thermal states, and the second is the implementation of controlled dynamical evolution. NMR systems are ensemble-based and naturally begin in a thermal equilibrium state \(\rho_{\mathrm{eq}} \sim \gamma_{\rm F}\sum_{i=1}^3\sigma_z^{i}+\gamma_{\rm H}\sum_{i=4}^5\sigma_z^{i}\), after ignoring the identity component for convenience, where $\gamma_{\rm F}$ and $\gamma_{\rm H}$ denote the gyromagnetic ratios of the fluorine and hydrogen nuclei, respectively. Starting from the initial state $\rho_{\mathrm{eq}}$, we can redistribute  the level populations via coherent control and subsequently eliminate the off-diagonal elements to arrive at a state close to the target $\rho_{0}(\beta) \sim \sigma_z \otimes \sum_{n=1}^{16} n^{-\beta} |n\rangle\langle n|$. 

 \begin{figure*}[hbtp]
     \begin{center}
     \includegraphics[width=2.1\columnwidth]{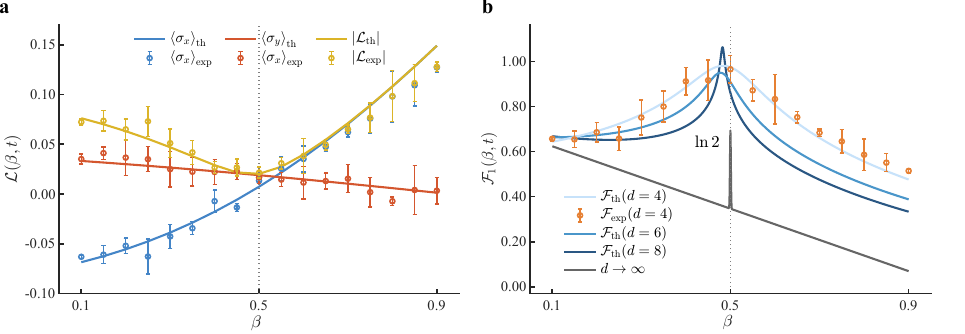}
     \end{center}
     \vspace{-0.50cm}
     \caption{\textbf{The coherence dynamics ${\cal L}(\beta, t)$  of the probe qubit  and the free energy density $ \mathcal{F}_1(\beta,t)$ of the quantum system 
     varying with the value of $\beta$.}  The evolution time $t$ is set to $14.13$, the imaginary part of the first RH zero. \textbf{a}, As $\beta$ passes through the point associated with the RH, the coherence approximately drops to zero, with its non-vanishing value being a consequence of finite-size effects. \textbf{b}, Non-analytic behavior of the free energy density  as $\beta$ passes through the point associated with the RH. In the thermodynamic limit $d \rightarrow \infty$, the free-energy density undergoes a sudden change from $ (1-\beta)\ln 2$ to $ \ln 2$ at $\beta=1/2$, signaling the occurrence of DQPTs.
     } 
     \vspace{-0.50cm}
     \label{Fig4}
 \end{figure*}

As shown in Fig.~\ref{Fig2}c, our procedure for preparing the thermal state primarily consists of the following steps. (i) Elimination of populations on all levels except the \(\ket{\uparrow\uparrow\uparrow\uparrow\uparrow}\) and \(\ket{\downarrow\downarrow\downarrow\downarrow\downarrow}\) levels. By exploiting the antisymmetric population distribution between pairs of energy levels in $\rho_{\mathrm{eq}}$, we apply selective $\pi/2$ rotations to each such pair, excluding the states  \(\ket{\uparrow\uparrow\uparrow\uparrow\uparrow}\) and \(\ket{\downarrow\downarrow\downarrow\downarrow\downarrow}\), to convert the diagonal population matrix into a purely off-diagonal form. 
Subsequently, a gradient field along the $z$-direction is applied to decohere these off-diagonal elements. 
(ii) Transfer of population from  \(\ket{\uparrow\uparrow\uparrow\uparrow\uparrow}\) and \(\ket{\downarrow\downarrow\downarrow\downarrow\downarrow}\)  to other levels, according to the target distribution encoded in $\rho_{0}(\beta)$. This is achieved by implementing coherent $R_y(\lambda)$ [or $R_y(\psi)$] rotations between \(\ket{\uparrow\uparrow\uparrow\uparrow}\) (or \(\ket{\downarrow\downarrow\downarrow\downarrow}\)) and other levels within the $\ket{\uparrow}$ [or $\ket{\downarrow}$] subspace of the probe qubit. The rotation angles are determined by $\beta$. During this process, the off-diagonal terms generated are also eliminated by applying a gradient pulse. However, since the gradient field alone cannot fully suppress coherence within homonuclear subspaces. SWAP operations are applied between the F and H spin subspaces in conjunction with the gradient field to ensure complete decoherence. Finally, the desired thermal state is prepared. Further details can be found in the Methods.\par

As outlined in Section~\ref{model1}, two controlled Hamiltonian dynamics under $\mathcal{H}_{\rm c1}$ and $\mathcal{H}_{\rm c2}$  are to be implemented. The first $\text{e}^{-i\mathcal{H}_{\rm c1} \pi/\lambda}$ corresponds to a controlled-$Z$ gate between the probe spin and last spin. The second $\text{e}^{-i\mathcal{H}_{\rm c2} t}$ can be decomposed into multiple  multi-qubit controlled-$R_z(\theta_n)$ operations. The rotation angles are given by $\theta_n=t\log n$, where $t$ denotes the imagnary part of the RH complex plane. Last, we measure the coherence $\mathcal{L}(\beta,t)=\bra{\uparrow}\rho_f\ket{\downarrow}$ of the probe spin on the finial state $\rho_f=U_{\rm tot}\rho_0U^\dagger_{\rm tot}$ where $U_{\rm tot}=\text{e}^{-i\mathcal{H}_{\rm c2} t}\text{e}^{-i\mathcal{H}_{\rm c1} \pi/\lambda}(H\otimes I)$. Figure \ref{Fig2} illustrates the full experimental sequence, from initial state preparation to final measurement. To enhance control accuracy, we implement all coherent operations and unitary evolutions using shaped pulses optimized via a gradient ascent pulse technique~\cite{exp4}, yielding a numerical simulation fidelity exceeding 99.5\%.

We conducted three types of experiments to comprehensively validate the theoretical predictions. First, we prepare the thermal state $\rho_0(\beta)$ with $\beta=0.5$ (corresponding to the RH) and vary the evolution time 
$t$ under the Hamiltonian $\mathcal{H}_{\rm c2}$. Second, we prepare another thermal state $\rho_0(\beta)$ with $\beta=0.3$, which lies off the critical line of the RH, we again vary the evolution time $t$. Third, we fix the evolution time at $t=14.13$, corresponding to the imaginary part of the first nontrivial Riemann zero, and scan $\beta$ over the range $[0.1, 0.9]$. Then the coherence dynamics ${\cal L}(\beta, t)$ of the probe qubit is measured in each configuration. Figures~\ref{Fig3}a and~\ref{Fig3}b present the coherence dynamics ${\cal L}(\beta, t)$ under different values of $\beta$. When $\beta=0.5$, the coherence of the probe spin exhibits recurring vanishing-and-revival behavior, and the points where both the real and imaginary parts of ${\cal L}(\beta, t)$ approach zero correspond to Riemann zeros. As a demonstration, $t_1=14.12$ ($t_1^{\rm th}=14.13$), $t_2=20.96$ ($t_2^{\rm th}=21.02$), $t_3=25.09$ ($t_3^{\rm th}=25.01$), $t_4=30.44$ ($t_4^{\rm th}=30.43$), and $t_5=32.93$ ($t_5^{\rm th}=32.94$) coherence zeros are extracted through polynomial fitting  of the experimental data. These values agree well with the ideal zeros of the zeta function. In contrast, for $\beta=0.3$, the coherence dynamics show no discernible vanishing-and-revival behavior, and no significant coherence zeros are observed. Figure~\ref{Fig4} shows the behavior of the coherence as a function of $\beta$. The results clearly indicate the emergence of coherence zeros  and  DQPTs  when crossing the critical line $\beta=0.5$, in accordance with the RH.

\subsection{Simulating the RH within a universal quantum computing framework}\label{QC}

Since the two quantum systems introduced above are strongly correlated, their direct realization on a gate-based quantum computer is highly nontrivial~\cite{efficent2025}.  Therefore, we propose an efficient computational framework that enables their realization on a universal gate-based quantum computer. The constructions  in Sections~\ref{model1} and~\ref{model2} can be reduced to two fundamental tasks: 
\begin{itemize}
    \item[(i)] Preparation of the initial thermal state, which encodes the real part 
    of the  zeta argument.
    \item[(ii)] Simulation of time evolution under the target Hamiltonian, which encodes the imaginary part.
\end{itemize}
 The  initial state  $|\psi_0\rangle=\tfrac{1}{C}\textstyle\sum_{n=1}^{N}n^{-\beta/2}|n\rangle$ can be prepared to precision $\varepsilon$ with success possibility at least $(\frac{1}{2}-\tfrac{\varepsilon}{3})$ using polynomial qubits and basic gates.  We further show that the evolution $U(t)=\mathrm{e}^{-i\mathcal{H}_0t}$ for $\mathcal{H}_0=\textstyle\sum_{n=1}^N \log n|n\rangle\langle n|$ can be also implemented to precision $\xi$ with polynomial resources, and controlled evolutions incur the same polynomial overhead. These guarantees ensure scalability of our method on a gate-based quantum computer~(see Methods~\ref{Methods:Construction} for details).\par
 We then apply this framework to accelerate the verification and probing of nontrivial zeros via the Riemann-Siegel formula, which decomposes the zeta function into two partial summations that can be evaluated  using systems similar to that in Section \ref{model1}. To achieve a total error $\delta$ in the zeta function,  the quantum circuit complexities are bounded by $\mathrm{Poly}(\log \delta^{-1},\log |t|,\log(1-\beta)^{-1},\log \beta^{-1})$, while the corresponding sample complexities are bounded by $\mathcal{O}(\delta^{-1})(1-\beta)^{-1}|t|^{(1-\beta)/2}$ and $\mathcal{O}(\delta^{-1})\beta^{-1}|t|^{(1-\beta)/2}$, respectively. Hence the overall computational complexity~(See Methods~\ref{Methods:complexity} for details) is bounded by:
\begin{equation}
    \delta^{-1}|t|^{(1-\beta)/2}
    \mathrm{Poly}(\log \delta^{-1},\log |t|,(1-\beta)^{-1},\beta^{-1}).
\end{equation}
Specializing to locating and verifying nontrivial zeros, where $ \tfrac{1}{2}\leq \beta \leq 1-\frac{1}{\mathcal{O} (\, (\log |t|)^{2/3} (\log \log |t|)^{1/3})}$~\cite{mossinghoff2024explicit}, this simplifies to
\begin{equation}
    \delta^{-1}|t|^{(1-\beta)/2}
    \mathrm{Poly}(\log \delta^{-1},\log |t|).
\end{equation} 
Thus the $|t|$-dependence reduces from $\sqrt{|t|}$ (direct Riemann–Siegel evaluation) to $|t|^{(1-\beta)/2}$ (at most $|t|^{1/4}$), yielding at least a quadratic speedup. Extended Data Table 1 summarizes these complexity bounds.

\section{Discussion}\label{dis}
In this work, we have established a direct correspondence between the RH and the dynamics of quantum many-body systems, and experimentally verified this connection on a physical platform. The mapping between nontrivial Riemann zeros and DQPTs reveals a deep and unexplored bridge between number theory and quantum physics. In particular, we have demonstrated that a thermally equilibrated quantum system, when driven by a tailored Hamiltonian, undergoes time-domain phase transitions whose critical points align precisely with the Riemann zeros. To verify large Riemann zeros, we also construct a quantum system  which undergoes time-domain phase transitions whose critical points align increasingly accurately to the large Riemann zeros.  
Together, these results suggest that the mysterious structure of the zeta function may admit a physical interpretation in terms of quantum critical dynamics.

Importantly, we demonstrated that both classes of systems considered here can be constructed on a universal quantum computer in polynomial time and  numerical verification of the RH can be achieved by our method with at least a quadratic speedup over classical methods. Beyond this, our results suggest that Riemann zeros also encode signatures of information propagation in quantum systems, motivating future studies of their relationship to out-of-time-ordered correlators (OTOCs)~\cite{garttner2017measuring,braumuller2022probing,google2025observation} and other diagnostics of quantum chaos.

Given the inherent difficulty of verifying the RH using classical computation, our approach opens a path toward leveraging quantum computing as an algorithmic tool for probing one of mathematics’ most enduring open problems. We note that more efficient classical methods for evaluating the Riemann zeta function on the critical line $\beta=\tfrac{1}{2}$ are known~\cite{odlyzko200110,hiary2011fast}. Our ongoing work explores how the proposed framework can accelerate the most time-consuming subroutines of these methods, thereby further improving their overall efficiency. Moreover, the framework presented here can be extended to the evaluation of other series and special functions, potentially defining a broader class of number-theoretic benchmarks for demonstrating quantum advantage. For example, the prime number distribution can be constructed using the Riemann function, thus allowing this computational framework to be further applied to the calculation of Witten index in string theory and supersymmetry theory.
\clearpage
\citestyle{nature}
\bibliography{RH.bib}

@article{srednicki2011nonclassical,
  title={{N}onclassical degrees of freedom in the {R}iemann {H}amiltonian},
  author={Srednicki, Mark},
  journal={Phys. Rev. Lett.},
  volume={107},
  number={10},
  pages={100201},
  year={2011},
  publisher={APS},
url={https://journals.aps.org/prl/abstract/10.1103/PhysRevLett.107.100201}
}

@article{bender2017hamiltonian,
  title={{H}amiltonian for the zeros of the {R}iemann zeta function},
  author={Bender, Carl M and Brody, Dorje C and M{\"u}ller, Markus P},
  journal={Phys. Rev. Lett.},
  volume={118},
  number={13},
  pages={130201},
  year={2017},
  publisher={APS},
url={https://journals.aps.org/prl/abstract/10.1103/PhysRevLett.118.130201}
}

@article{schumayer2011colloquium,
  title={{C}olloquium: {P}hysics of the {R}iemann hypothesis},
  author={Schumayer, D{\'a}niel and Hutchinson, David AW},
  journal={Rev. Mod. Phys.},
  volume={83},
  number={2},
  pages={307--330},
  year={2011},
  publisher={APS},
url={https://journals.aps.org/rmp/abstract/10.1103/RevModPhys.83.307}
}

@article{gutzwiller1983stochastic,
  title={{S}tochastic behavior in quantum scattering},
  author={Gutzwiller, Martin C},
  journal={Phys. D (Amsterdam, Neth.)},
  volume={7},
  number={1-3},
  pages={341--355},
  year={1983},
  publisher={Elsevier},
url={https://www.sciencedirect.com/science/article/abs/pii/0167278983901380}
}

@article{bhaduri1995phase,
  title={{P}hase of the {R}iemann $\zeta$ function and the inverted harmonic oscillator},
  author={Bhaduri, RK and Khare, Avinash and Law, J},
  journal={	Phys. Rev. E},
  volume={52},
  number={1},
  pages={486},
  year={1995},
  publisher={APS},
url={https://journals.aps.org/pre/abstract/10.1103/PhysRevE.52.486}
}

@article{peng2015experimental,
  title={{E}xperimental observation of {Lee-Yang} zeros},
  author={Peng, Xinhua and Zhou, Hui and Wei, Bo-Bo and Cui, Jiangyu and Du, Jiangfeng and Liu, Ren-Bao},
  journal={Phys. Rev. Lett.},
  volume={114},
  number={1},
  pages={010601},
  year={2015},
  publisher={APS},
url={https://journals.aps.org/prl/abstract/10.1103/PhysRevLett.114.010601}
}

@article{wei2012lee,
  title={{L}ee-{Y}ang zeros and critical times in decoherence of a probe spin coupled to a bath},
  author={Wei, Bo-Bo and Liu, Ren-Bao},
  journal={Phys. Rev. Lett.},
  volume={109},
  number={18},
  pages={185701},
  year={2012},
  publisher={APS},
url={https://journals.aps.org/prl/abstract/10.1103/PhysRevLett.109.185701}
}

@inproceedings{borwein2000efficient,
  title={{A}n efficient algorithm for the {R}iemann zeta function},
  author={Borwein, Peter},
  booktitle={{C}anadian {M}athematical {S}ociety {C}onference {P}roceedings},
  volume={27},
  pages={29--34},
  year={2000},
url={https://www.cecm.sfu.ca/~pborwein/PAPERS/P155.pdf}
}

@article{odlyzko200110,
  title={{T}he $10^{22}$-nd zero of the {R}iemann zeta function},
  author={Odlyzko, Andrew M},
  journal={Contemp. Math.},
  volume={290},
  pages={139--144},
  year={2001},
  publisher={Providence, RI; American Mathematical Society; 1999},
url={https://www-users.cse.umn.edu/~odlyzko/doc/zeta.10to22.ps}
}

@article{bui2011large,
  title={{L}arge gaps between consecutive zeros of the {R}iemann zeta-function},
  author={Bui, HM},
  journal={J. Number Theory},
  volume={131},
  number={1},
  pages={67--95},
  year={2011},
  publisher={Elsevier},
url={https://www.sciencedirect.com/science/article/pii/S0022314X10002131}
}

@inproceedings{berry2005riemann,
  title={{R}iemann's zeta function: {A} model for quantum chaos?},
  author={Berry, Michael V},
  booktitle={{Q}uantum {C}haos and {S}tatistical {N}uclear {P}hysics: {P}roceedings of the 2nd {I}nternational {C}onference on {Q}uantum {C}haos and the 4th {I}nternational {C}olloquium on {S}tatistical {N}uclear {P}hysics, {H}eld at {C}uernavaca, {M}{\'e}xico, {J}anuary 6--10, 1986},
  pages={1--17},
  year={2005},
  organization={Springer},
url={https://link.springer.com/chapter/10.1007/3-540-17171-1_1}
}

@article{berry1992new,
  title={{A} new asymptotic representation for $\zeta$ ($1/2$+ it) and quantum spectral determinants},
  author={Berry, Michael Victor and Keating, Jonathan P},
  journal={Proc. R. Soc. Lond., Ser. A},
  volume={437},
  number={1899},
  pages={151--173},
  year={1992},
  publisher={The Royal Society London},
url={https://royalsocietypublishing.org/doi/abs/10.1098/rspa.1992.0053}
}

@article{remmen2021amplitudes,
  title={{A}mplitudes and the {R}iemann {Z}eta function},
  author={Remmen, Grant N},
  journal={Phys. Rev. Lett.},
  volume={127},
  number={24},
  pages={241602},
  year={2021},
  publisher={APS},
url={https://journals.aps.org/prl/abstract/10.1103/PhysRevLett.127.241602}
}

@article{heyl2013dynamical,
  title={{D}ynamical quantum phase transitions in the transverse-field {I}sing model},
  author={Heyl, Markus and Polkovnikov, Anatoli and Kehrein, Stefan},
  journal={Phys. Rev. Lett.},
  volume={110},
  number={13},
  pages={135704},
  year={2013},
  publisher={APS},
url={https://journals.aps.org/prl/abstract/10.1103/PhysRevLett.110.135704}
}

@article{karrasch2013dynamical,
  title={{D}ynamical phase transitions after quenches in nonintegrable models},
  author={Karrasch, C and Schuricht, D},
  journal={Phys. Rev. B},
  volume={87},
  number={19},
  pages={195104},
  year={2013},
  publisher = {APS},
url={https://journals.aps.org/prb/abstract/10.1103/PhysRevB.87.195104}
}

@article{heyl2018dynamical,
  title={{D}ynamical quantum phase transitions: a review},
  author={Heyl, Markus},
  journal={Rep. Progr. Phys.},
  volume={81},
  number={5},
  pages={054001},
  year={2018},
  publisher={IOP Publishing},
url={https://iopscience.iop.org/article/10.1088/1361-6633/aaaf9a/meta}
}

@article{long2001efficient,
  title={{E}fficient scheme for initializing a quantum register with an arbitrary superposed state},
  author={Long, Gui-Lu and Sun, Yang},
  journal={Phys. Rev. A},
  volume={64},
  number={1},
  pages={014303},
  year={2001},
  publisher={APS},
url={https://journals.aps.org/pra/abstract/10.1103/PhysRevA.64.014303}
}

@article{grover2002creating,
  title={{C}reating superpositions that correspond to efficiently integrable probability distributions},
  author={Grover, Lov and Rudolph, Terry},
  journal={arXiv preprint quant-ph/0208112},
  year={2002},
url={https://arxiv.org/abs/quant-ph/0208112}
}

@article{akiyama2001multiple,
  title={{M}ultiple zeta values at non-positive integers},
  author={Akiyama, Shigeki and Tanigawa, Yoshio},
  journal={The Ramanujan Journal},
  volume={5},
  number={4},
  pages={327--351},
  year={2001},
  publisher={Springer},
url={https://link.springer.com/article/10.1023/A:1013981102941}
}

@article{leclair1995boundary,
  title={{B}oundary energy and boundary states in integrable quantum field theories},
  author={LeClair, Andr{\'e} and Mussardo, Giuseppe and Saleur, H and Skorik, S},
  journal={Nucl. Phys. B},
  volume={453},
  number={3},
  pages={581--618},
  year={1995},
  publisher={Elsevier},
url={https://www.sciencedirect.com/science/article/pii/055032139500435U}
}

@article{litinski2024quantum,
  title={{Q}uantum schoolbook multiplication with fewer {T}offoli gates},
  author={Litinski, Daniel},
  journal={arXiv preprint arXiv:2410.00899},
  year={2024},
url={https://arxiv.org/abs/2410.00899}
}

@article{mossinghoff2024explicit,
  title={{E}xplicit zero-free regions for the {R}iemann zeta-function},
  author={Mossinghoff, Michael J and Trudgian, Timothy S and Yang, Andrew},
  journal={Res. Number Theory},
  volume={10},
  number={1},
  pages={11},
  year={2024},
  publisher={Springer},
url={https://link.springer.com/article/10.1007/s40993-023-00498-y}
}

@article{hiary2011fast,
  title={{F}ast methods to compute the {R}iemann zeta function},
  author={Hiary, Ghaith Ayesh},
  journal={Annals of Math.},
  pages={891--946},
  year={2011},
  publisher={JSTOR},
url={https://www.jstor.org/stable/23030516}
}

@article{heyl2017dynamical,
  title={{D}ynamical topological quantum phase transitions for mixed states},
  author={Heyl, Markus and Budich, JC},
  journal={Phys. Rev. B},
  volume={96},
  number={18},
  pages={180304},
  year={2017},
  publisher={APS},
url={https://journals.aps.org/prb/abstract/10.1103/PhysRevB.96.180304}
}

@article{bhattacharya2017mixed,
  title={{M}ixed state dynamical quantum phase transitions},
  author={Bhattacharya, Utso and Bandyopadhyay, Souvik and Dutta, Amit},
  journal={Phys. Rev. B},
  volume={96},
  number={18},
  pages={180303},
  year={2017},
  publisher={APS},
url={https://journals.aps.org/prb/abstract/10.1103/PhysRevB.96.180303}
}

@article{lang2018dynamical,
  title={{D}ynamical quantum phase transition for mixed states in open systems},
  author={Lang, Haifeng and Chen, Yixin and Hong, Qiantan and Fan, Heng},
  journal={Phys. Rev. B},
  volume={98},
  number={13},
  pages={134310},
  year={2018},
  publisher={APS},
url={https://journals.aps.org/prb/abstract/10.1103/PhysRevB.98.134310}
}

@article{francis2021many,
  title={{M}any-body thermodynamics on quantum computers via partition function zeros},
  author={Francis, Akhil and Zhu, Daiwei and Huerta Alderete, Cinthia and Johri, Sonika and Xiao, Xiao and Freericks, James K and Monroe, Christopher and Linke, Norbert M and Kemper, Alexander F},
  journal={	Sci. Adv.},
  volume={7},
  number={34},
  pages={eabf2447},
  year={2021},
  publisher={American Association for the Advancement of Science},
url={https://www.science.org/doi/full/10.1126/sciadv.abf2447}
}

@article{exp1,
  title={{Quantum simulations of a particle in one-dimensional potentials using NMR}},
  author={Shankar, Ravi and Hegde, Swathi S and Mahesh, TS},
  journal={Phys. Lett. A},
  volume={378},
  number={1-2},
  pages={10--15},
  year={2014},
  publisher={Elsevier},
 url={https://doi.org/10.1016/j.physleta.2013.10.029}
}

@article{exp2,
  title={{Experimentally probing topological order and its breakdown through modular matrices}},
  author={Luo, Zhihuang and Li, Jun and Li, Zhaokai and Hung, Ling-Yan and Wan, Yidun and Peng, Xinhua and Du, Jiangfeng},
  journal={Nat. Phys.},
  volume={14},
  number={2},
  pages={160--165},
  year={2018},
  publisher={Nature Publishing Group UK London},
 url={https://doi.org/10.1038/nphys4281}
}

@article{exp3,
  title={{S}elf-{C}onsistent {D}etermination of {S}ingle-{I}mpurity {A}nderson {M}odel {U}sing {H}ybrid {Q}uantum-{C}lassical {A}pproach on a {S}pin {Q}uantum {S}imulator},
  author={Nie, Xinfang and Zhu, Xuanran and Fan, Yu-ang and Long, Xinyue and Liu, Hongfeng and Huang, Keyi and Xi, Cheng and Che, Liangyu and Zheng, Yuxuan and Feng, Yufang and others},
  journal={Phys. Rev. Lett.},
  volume={133},
  number={14},
  pages={140602},
  year={2024},
  publisher={APS},
  url={https://doi.org/10.1103/PhysRevLett.133.140602}
}

@article{exp4,
  title={{Optimal control of coupled spin dynamics: design of NMR pulse sequences by gradient ascent algorithms}},
  author={Khaneja, Navin and Reiss, Timo and Kehlet, Cindie and Schulte-Herbr{\"u}ggen, Thomas and Glaser, Steffen J},
  journal={J. Magn. Reson.},
  volume={172},
  number={2},
  pages={296--305},
  year={2005},
  publisher={Elsevier},
  url={https://doi.org/10.1016/j.jmr.2004.11.004}
}

@article{milgram2013integral,
  title={{I}ntegral and {S}eries {R}epresentations of {R}iemann' s {Z}eta {F}unction and {D}irichlet' s {E}ta {F}unction and a {M}edley of {R}elated {R}esults},
  author={Milgram, Michael S},
  journal={J. Math.},
  volume={2013},
  number={1},
  pages={181724},
  year={2013},
  publisher={Wiley Online Library},
    url={https://onlinelibrary.wiley.com/doi/pdf/10.1155/2013/181724}
}

@article{hughes2000random,
  title={{R}andom matrix theory and the derivative of the {R}iemann zeta function},
  author={Hughes, Chris P and Keating, Jon P and 'Connell, Neil},
  journal={Proc. R. Soc. Lond., Ser. A},
  volume={456},
  number={2003},
  pages={2611--2627},
  year={2000},
  publisher={The Royal Society},
    url={https://royalsocietypublishing.org/doi/epdf/10.1098/rspa.2000.0628}
}

@article{de2011high,
  title={{H}igh precision computation of {R}iemann’s zeta function by the {R}iemann-{S}iegel formula, {I}},
  author={de Reyna, J},
  journal={Math. Comput.},
  volume={80},
  number={274},
  pages={995--1009},
  year={2011},
url={https://www.ams.org/journals/mcom/2011-80-274/S0025-5718-2010-02426-3/S0025-5718-2010-02426-3.pdf}
}

@article{Bober03042018,
author = {Jonathan W. Bober and Ghaith A. Hiary},
title = {{N}ew {C}omputations of the {R}iemann {Z}eta {F}unction on the {C}ritical {L}ine},
journal = {Exp. Math.},
volume = {27},
number = {2},
pages = {125--137},
year = {2018},
publisher = {Taylor \& Francis},
doi = {10.1080/10586458.2016.1233083},
URL = { 
https://doi.org/10.1080/10586458.2016.1233083
}
}

@article{braumuller2022probing,
  title={{P}robing quantum information propagation with out-of-time-ordered correlators},
  author={Braum{\"u}ller, Jochen and Karamlou, Amir H and Yanay, Yariv and Kannan, Bharath and Kim, David and Kjaergaard, Morten and Melville, Alexander and Niedzielski, Bethany M and Sung, Youngkyu and Veps{\"a}l{\"a}inen, Antti and others},
  journal={Nat. Phys.},
  volume={18},
  number={2},
  pages={172--178},
  year={2022},
  publisher={Nature Publishing Group UK London},
URL = { 
https://www.nature.com/articles/s41567-021-01430-w
}
}

@article{garttner2017measuring,
  title={{M}easuring out-of-time-order correlations and multiple quantum spectra in a trapped-ion quantum magnet},
  author={G{\"a}rttner, Martin and Bohnet, Justin G and Safavi-Naini, Arghavan and Wall, Michael L and Bollinger, John J and Rey, Ana Maria},
  journal={Nat. Phys.},
  volume={13},
  number={8},
  pages={781--786},
  year={2017},
  publisher={Nature Publishing Group UK London},
URL = { 
https://www.nature.com/articles/nphys4119
}
}

@article{muniz2020exploring,
  title={{E}xploring dynamical phase transitions with cold atoms in an optical cavity},
  author={Muniz, Juan A and Barberena, Diego and Lewis-Swan, Robert J and Young, Dylan J and Cline, Julia RK and Rey, Ana Maria and Thompson, James K},
  journal={Nature},
  volume={580},
  number={7805},
  pages={602--607},
  year={2020},
  publisher={Nature Publishing Group UK London},
URL = { 
https://www.nature.com/articles/s41586-020-2224-x
}
}

@article{mack2010riemann,
  title={{R}iemann $\zeta$ function from wave-packet dynamics},
  author={Mack, R{\"u}diger and Dahl, Jens Peder and Moya-Cessa, Hector and Strunz, Walter T and Walser, Reinhold and Schleich, Wolfgang P},
  journal={Phys. Rev. A},
  volume={82},
  number={3},
  pages={032119},
  year={2010},
  publisher={APS},
URL = { 
https://doi.org/10.1103/PhysRevA.82.032119
        }
}

@article{feiler2013entanglement,
  title={{E}ntanglement and analytical continuation: an intimate relation told by the {R}iemann zeta function},
  author={Feiler, Cornelia and Schleich, Wolfgang P},
  journal={New J. Phys.},
  volume={15},
  number={6},
  pages={063009},
  year={2013},
  publisher={IOP Publishing},
url={https://iopscience.iop.org/article/10.1088/1367-2630/15/6/063009/meta}
}

@article{efficent2025,
	date = {2025/10/01},
    	author = {Chen, Chi-Fang and Kastoryano, Michael and Brand{\~a}o, Fernando G. S. L. and Gily{\'e}n, Andr{\'a}s},
	doi = {10.1038/s41586-025-09583-x},
	id = {Chen2025},
	isbn = {1476-4687},
	journal = {Nature},
	number = {8085},
	pages = {561--566},
	title = {{E}fficient quantum thermal simulation},
	url = {https://doi.org/10.1038/s41586-025-09583-x},
	volume = {646},
	year = {2025},
}

@article{google2025observation,
  title={Observation of constructive interference at the edge of quantum ergodicity},
author={{Google Quantum AI and Collaborators}},
  journal={Nature},
  volume={646},
  number={8086},
  pages={825--830},
  year={2025},
  publisher={Nature Publishing Group UK London},
url={https://www.nature.com/articles/s41586-025-09526-6}
}

@article{zhou2018dynamical,
  title={Dynamical quantum phase transitions in non-{H}ermitian lattices},
  author={Zhou, Longwen and Wang, Qing-hai and Wang, Hailong and Gong, Jiangbin},
  journal={Phys. Rev. A},
  volume={98},
  number={2},
  pages={022129},
  year={2018},
  publisher={APS},
url={https://doi.org/10.1103/PhysRevA.98.022129}
}

@article{spector1990supersymmetry,
  title={Supersymmetry and the {M{\"o}bius} inversion function},
  author={Spector, Donald},
  journal={Commun.Math. Phys.},
  volume={127},
  number={2},
  pages={239--252},
  year={1990},
  publisher={Springer},
url={https://link.springer.com/article/10.1007/BF02096755}
}

@article{cassettari2023holographic,
  title={Holographic realization of the prime number quantum potential},
  author={Cassettari, Donatella and Mussardo, Giuseppe and Trombettoni, Andrea},
  journal={PNAS },
  volume={2},
  number={1},
  pages={pgac279},
  year={2023},
  publisher={Oxford University Press},
url={https://academic.oup.com/pnasnexus/article/2/1/pgac279/6888012}
}

@article{gleisberg2013factorization,
  title={Factorization with a logarithmic energy spectrum},
  author={Gleisberg, Ferdinand and Mack, R{\"u}diger and Vogel, Karl and Schleich, Wolfgang P},
  journal={New J. Phys.},
  volume={15},
  number={2},
  pages={023037},
  year={2013},
  publisher={IOP Publishing},
url={https://iopscience.iop.org/article/10.1088/1367-2630/15/2/023037/meta}
}

@article{feiler2015dirichlet,
  title={Dirichlet series as interfering probability amplitudes for quantum measurements},
  author={Feiler, Cornelia and Schleich, Wolfgang P},
  journal={New J. Phys.},
  volume={17},
  number={6},
  pages={063040},
  year={2015},
  publisher={IOP Publishing},
url={https://iopscience.iop.org/article/10.1088/1367-2630/17/6/063040}
}

 \clearpage

\section{Methods}
\subsection{Thermal state preparation in experiment}\label{Methods:Thermal state}
The preparation sequence, shown in the main text, consists of two analytically designed line-selective shaped pulses, four pulsed field gradients, and three SWAP gates. The first shaped pulse selectively depletes population from all states except $\ket{\uparrow\uparrow\uparrow\uparrow\uparrow}$ and $\ket{\downarrow\downarrow\downarrow\downarrow\downarrow}$, which serve as logical sources for encoding the thermal-equivalent distribution. A gradient pulse is applied immediately afterward to suppress non-diagonal coherence terms generated in this step. The second shaped pulse redistributes the populations from $\ket{\uparrow\uparrow\uparrow\uparrow\uparrow}$ and $\ket{\downarrow\downarrow\downarrow\downarrow\downarrow}$ into a target position matching the desired thermal distribution. This step introduces zeroth-order coherences that cannot be fully removed by gradients alone. To address this, three SWAP operations are inserted before the final two gradient pulses, transferring coherence terms from homonuclear fluorine spins to heteronuclear channels, where they can be completely dephased. Below, we detail the function of each component in the preparation protocol.

The thermal equilibrium state of this five-qubit system is highly mixed. It can be described by a deviation density matrix of the form:
\begin{equation}
\rho_{\mathrm{eq}} \propto \gamma_F(\sigma_z^1 + \sigma_z^2 + \sigma_z^3) + \gamma_H(\sigma_z^4 + \sigma_z^5),
\end{equation}
where $\gamma_H$ and $\gamma_F$ are the gyromagnetic ratios for $^{1}\mathrm{H}$ and $^{19}\mathrm{F}$ nuclei, and $\sigma_z^i$ is the Pauli-$Z$ operator for the $i$-th qubit. This matrix is traceless and diagonal, with the maximal populations residing in the $\ket{\uparrow\uparrow\uparrow\uparrow\uparrow}$ and $\ket{\downarrow\downarrow\downarrow\downarrow\downarrow}$ states. Furthermore, pairs of symmetric basis states (e.g., $\ket{\uparrow\downarrow\uparrow\downarrow\uparrow}$ and $\ket{\downarrow\uparrow\downarrow\uparrow\downarrow}$) possess populations of equal magnitude but opposite sign, satisfying $P_{\ket{\uparrow\downarrow\uparrow\downarrow\uparrow}} + P_{\ket{\downarrow\uparrow\downarrow\uparrow\downarrow}} = 0$. This intrinsic symmetry is key to our line-selective approach, allowing us to manipulate populations.

The first shaped pulse is designed to selectively deplete the population of all energy levels except for the $\ket{\uparrow\uparrow\uparrow\uparrow\uparrow}$ and $\ket{\downarrow\downarrow\downarrow\downarrow\downarrow}$ states, which serve as the source states for subsequent redistribution. This is achieved by applying a series of analytically constructed $\pi/2$ line-selective rotations between all pairs of symmetric basis states(e.g., $\ket{\uparrow\downarrow\uparrow\downarrow\uparrow}$ and $\ket{\downarrow\uparrow\downarrow\uparrow\downarrow}$). For a given pair of states, say $\ket{x}$ and $\ket{\Bar{x}}$, their initial populations are represented by a diagonal matrix $\begin{bmatrix} a & 0 \\ 0 & -a \end{bmatrix}$. Applying an $R_y(\pi/2)$ rotation transforms this to $\begin{bmatrix} 0 & a \\ a & 0 \end{bmatrix}$. Since these pairs have opposite initial populations in $\rho_{\mathrm{eq}}$, applying a $\pi/2$ rotation along the $y$-axis effectively converts the diagonal population matrix to a purely off-diagonal one, which is then removed by a subsequent gradient pulse along the $z$-direction.
Because the system has an odd number of qubits, this procedure avoids generating zeroth-order coherences. A pulsed field gradient is then applied to remove any residual higher-order coherences, leading to a diagonal intermediate state $\rho_1$ in which only the $\ket{\uparrow\uparrow\uparrow\uparrow\uparrow}$ and $\ket{\downarrow\downarrow\downarrow\downarrow\downarrow}$ populations remain.

The second shaped pulse redistributes the populations of $\ket{\uparrow\uparrow\uparrow\uparrow\uparrow}$ and $\ket{\downarrow\downarrow\downarrow\downarrow\downarrow}$ into a coherent superposition that encodes the target thermal-equivalent state. This state is represented by:
\begin{equation}
\rho = \sigma_z \otimes \sum_{n=1}^{16} \frac{n^{-\beta}}{\mathcal{N}^2} |n\rangle \langle n|,
\end{equation}
with the normalization constant $\mathcal{N}$.
This is accomplished by a sequence of analytically determined line-selective rotations that couple the source basis to the remaining basis. Each rotation is precisely calculated to transfer a fraction of the population from $\ket{\uparrow\uparrow\uparrow\uparrow\uparrow}$ or $\ket{\downarrow\downarrow\downarrow\downarrow\downarrow}$ to a selected target basis. The rotation angles are derived from $\theta_k = 2\sin^{-1}\left(\sqrt{p_k / P_{\mathrm{rem}}}\right)$, where $p_k$ is the target population of the $k$-th basis and $P_{\mathrm{rem}}$ is the remaining unallocated population.
This {asymmetric redistribution process inevitably introduces zeroth-order coherences}, particularly among homonuclear fluorine spins, which cannot be effectively suppressed by gradient fields alone.

To address the issue of residual homonuclear coherences, we employ a {SWAP-assisted gradient scheme}. The core idea is to transfer these coherences from homonuclear channels to heteronuclear ones, where they can be completely dephased by a gradient pulse.
The sequence begins with a {SWAP(1,4)} gate to exchange the states of qubit 1 ($^{19}\mathrm{F}$) and qubit 4 ($^{1}\mathrm{H}$). This operation converts fluorine-fluorine coherences (e.g., between qubits 1 and 2) into heteronuclear coherences that can be dephased by the first gradient pulse. A second {SWAP(1,4)} gate restores the original logical labeling. To eliminate the remaining fluorine-fluorine coherence between qubits 2 and 3, we next apply a {SWAP(2,4)} gate, shifting this coherence to the heteronuclear (3,4) channel, which is then removed by the second gradient pulse.
For increased efficiency, the two middle SWAP gates, {SWAP(1,4)} and {SWAP(2,4)}, can be combined into a single {SWAP(1,2)} operation. A final {SWAP(2,4)} then returns the system to its original qubit configuration. This sequence of SWAP gates and gradient pulses ensures that all zeroth-order coherences are effectively eliminated, yielding a clean thermal-equivalent state with a numerical fidelity exceeding 99\%.

\subsection{Non-analytic behavior of the free energy}\label{Methods:free energy}
The free energy density is defined as
\begin{equation}
    \mathcal{F}_1(\beta,t)=\lim_{N\rightarrow \infty}-\frac{1}{d}\ln |\mathcal{L}(\beta,t)|,
\end{equation}
where $d=\log N$ denotes the number of degrees of freedom. According to  Eq. \eqref{eq:L}, the accumulated phase factor can be expressed as
\begin{equation}
        {\cal L}(\beta,t)
    =S_N(s)/{\cal Z}(\beta,\mathcal{H}_0),
\end{equation}
 where $S_N(s)=-\textstyle\sum\limits_{n=1}^{N}(-1)^{n+1}n^{-s}$.
We prove that~(see Supplementary Notes 1 and 2):
\begin{equation}
    \textstyle\lim\limits_{N\rightarrow\infty}-\tfrac{1}{d}\ln(|S_N(s)|)
    =
    \begin{cases}
        \beta \ln2,\:&\text{Riemann zeros},\\
        0,\: &\text{Otherwise},
    \end{cases}
\end{equation}
and 
\begin{equation}
    \textstyle\lim\limits_{N\rightarrow\infty}-\tfrac{1}{d}\ln({\cal Z}(\beta,\mathcal{H}_0))
    =
    \begin{cases}
        (\beta-1)\ln2,\:&0<\beta<1,\\
        0,\: &\beta>1.
    \end{cases}
\end{equation} Combining these results, the free energy density in the thermodynamic limit is given by
\begin{align}\label{eq:thermal}
&\mathcal{F}_1(\beta,t)\notag\\
=&\lim_{N\rightarrow\infty}-\tfrac{1}{d}\Bigl(\ln(|
S_N(s)|)-\ln({\cal Z}(\beta,\mathcal{H}_0))\Bigr)\notag\\
    =&\begin{cases}
        \ln2, &\text{Riemann zeros,}\\
        (1-\beta)\ln 2, & 0<\beta<1 \text{, $\notin$ Riemann zeros,}\\
        0, &\beta>1.
    \end{cases}
\end{align}
This result demonstrates that the free energy density exhibits a non-analytic divergence at the points corresponding to the nontrivial Riemann zeros, providing strong evidence for the occurrence of  DQPTs.

\subsection{Measuring the Loschmidt amplitude}\label{Methods:Measure LA}
We consider a system, similar to the one described in Section~\ref{model2},  designed to measure the Loschmidt amplitude defined in Eq.~\eqref{Loschmidt amplitude}. The composite system is initialized in the same state $\rho_0 = \ket{\uparrow}\bra{\uparrow} \otimes \rho_s$, but now evolves  under the coupling Hamiltonian
$\mathcal{H}_{\rm c2} =\tfrac{1}{2} \sigma_x \otimes (\mathcal{H}_0-\dot{\theta}(t)\mathcal{I})$, which is  one half of the Hamiltonian $\mathcal{H}_{\rm c}$. The corresponding time-evolution operator is therefore $U(t) = \exp(-i\int_{0}^{t}\mathcal{H}_{\rm c2} dt')=\exp(-\tfrac{i}{2}\int_{0}^{t}\mathcal{H}_{\rm c} dt' )$. Defining $A(t) = H_0 t - \theta(t)$,  the  density matrix after evolution is
\begin{align}
    U \rho_0 U^\dagger = &\frac{1}{2} \Big[ \ket{+}\bra{+} \otimes \rho_s + \mathrm{e}^{-iA(t)} |+\rangle \langle -| \otimes \rho_s \notag\\
    +& \mathrm{e}^{iA(t)} |-\rangle \langle +| \otimes \rho_s + |-\rangle \langle -| \otimes \rho_s \Big],
\end{align}
in which $|+\rangle = \tfrac{1}{\sqrt{2}} (\ket{\uparrow} + \ket{\downarrow}) $ and $ |-\rangle=\tfrac{1}{\sqrt{2}} (\ket{\uparrow} - \ket{\downarrow})$. The reduced density matrix of the probe qubit is
\begin{align}
\rho_{\mathrm{probe}}=\frac{1}{2} I+\frac{1}{2}\Re S(t) \sigma_z+\frac{1}{2}\Im S(t) \sigma_y,
\end{align}
in which
\begin{equation}
    S(t) = \mathrm{Tr}[\mathrm{e}^{-iA(t)} \rho_s] 
= \tfrac{1}{\mathcal{Z}(\beta,H_0)} \mathrm{e}^{i\theta(t)} \textstyle\sum\limits_{n=1}^{N} n^{-1/2 - i t}.
\end{equation}
Consequently, measuring $\langle\sigma_z\rangle$ on the first qubit yields $\Re S(t)={\cal G}(\tfrac{1}{2},t)$.

\subsection{Construction of the quantum systems on  gate-based quantum computers}\label{Methods:Construction}
In this subsection, we describe the construction of the target initial state $|\psi_0\rangle=\tfrac{1}{C}\textstyle\sum_{n=1}^{N}n^{-\beta/2}|n\rangle$ and the evolution operator $U(t)=\mathrm{e}^{-i\mathcal{H}_0t}$ with $\mathcal{H}_0=\textstyle\sum_{n=1}^{N} \log(n)|n\rangle\langle n|$ on a standard gate-based quantum computer. We begin with two preliminary lemmas. The first establishes a quantum oracle for evaluating polynomial functions, while the second provides a procedure for approximating  logarithmic functions to required precision. Together, these results serve as building blocks for our main results: a  method for constructing the initial state (Theorem~\ref{th:init}) and an efficient implementation of the evolution operator (Theorem~\ref{Theorem:H}).\par

\begin{Lemma}[Polynomial oracle]\label{Le:poly}
Let $O(f)$ denote an oracle implementing the transformation 
\begin{equation}
    \sum_x\alpha_x|x\rangle|0\rangle\mapsto \sum_x\alpha_x|x\rangle|f(x)\rangle,
\end{equation}
where $f(x)$ is a polynomial of degree at most $D$. Suppose the coefficients of $f$ and the input $x$ are specified to $a_1$ and $a_2$ significant digits, respectively. The output is encoded with $r_1$ integer qubits and $r_2$ fractional qubits.
 Then  $O(f)$ can be implemented using $(D^{2}a_2^{2} + D^{2}a_1a_2 + D(r_1+r_2))$ gates and $(2Da_2+a_1)$ ancilla qubits.  
\end{Lemma}
\emph{Proof sketch.}  The oracle is implemented by sequentially constructing monomials $x^{d}$ for $d \in [0,D]$, multiplying them by coefficients $c_d$, and summing the results (see Supplementary Note 3 for details).

 \begin{Lemma}[Logarithm oracle]\label{le:log}
    Let $L$ denote an oracle implementing the transformation 
    \begin{equation}
        \sum_{n=1}^{N} \alpha_n|n\rangle|0\rangle \mapsto \sum_{n=1}^{N}\alpha_n|n\rangle|\widetilde{\log n}\rangle,
    \end{equation}
    where $\widetilde{\log n}$ approximates $\log n$ to within error $\eta$. Then $L$ can be implemented using 
       $\mathcal{O}((\log N)^3 \log^2(1/\eta))$
    gates and $\mathcal{O}((\log N)^2 \log(1/\eta))$ ancilla qubits. 
\end{Lemma}
\emph{Proof sketch.} The construction proceeds in two steps.
First, a register of ancilla qubits 
identifies the interval
\begin{equation}
    P_\nu:\; 3\cdot 2^{\nu-1} \leq n < 3\cdot 2^\nu,
    \qquad 1\leq \nu \leq \Bigl\lceil \log (N/3)\Bigr\rceil,
\end{equation}
yielding a contribution $(\nu+1)$ in $\log n$.
Second, approximate the residual contribution 
\begin{equation}
   \log(1+d), \qquad d = n/2^{\nu+1}-1,
\end{equation}
to accuracy $\eta$ using the polynomial oracle in Lemma~\ref{Le:poly}. 
The result is encoded using $\log N$ qubits for the integer part and $\log(1/\eta)$ for the fractional part. Full details of the polynomial expansions and resource counts are provided in  Supplementary Note 3.

\subsubsection{Initial state preparation}
The target initial state can be written (formally, up to normalization and coefficients) as
\begin{equation}
    |\psi_0\rangle=|\psi_0\rangle_1+|\psi_0\rangle_2- |\psi_0\rangle_3.
\end{equation}
We first construct of $|\psi_0\rangle_1$ using the Angle-preparation oracle $U_m$, and then  recover the initial state via the linear combination of Unitary (LCU) method and post-selection. A complete proof is provided in Supplementary Note 4.
\begin{Theorem}[Truncated state preparation]\label{th:tru}
    Let $\beta>0,\;\beta \neq 1$, and let $n_0,n_1\in\mathbb{N}$ with $n_1-n_0=2^k$ for some integer $k>0$. Define
    \begin{equation}
        |\psi_0 \rangle_1 =\frac{1}{C_1}\textstyle\sum\limits_{n=n_0}^{n_1-1} n^{-\beta/2}\,|n-n_0\rangle,
    \;
        C_1=\sqrt{ \textstyle\sum\limits_{n=n_0}^{n_1-1} n^{-\beta}}.
    \end{equation}
    Then  $|\psi_0\rangle_1$ can be prepared on a standard gate-based quantum computer to precision $\varepsilon>0$, using a number of gates and ancilla qubits bounded by 
    \begin{equation}
        \mathrm{Poly}(\log \varepsilon^{-1}, \log n_1,  \log|1-\beta|^{-1},  \beta,  v ),
    \end{equation}
    where $v$ denotes the number of significant digits used to represent $\beta$. The parameters are required to satisfy
    \begin{equation}
        c =\lceil \tfrac{1}{2}\log_{2\pi}(8\varepsilon^{-1})\rceil, 
        \qquad n_0 > \lceil \beta + 2c \rceil.
    \end{equation}
\end{Theorem}

\noindent
\emph{Proof sketch.}  
The construction adapts Grover's amplitude splitting method \cite{grover2002creating}, replacing integrals with partial summations.  The desired distribution is generated by $k$ iterative amplitude splits. At step $m$ ($0 \le m \le k-1$), each bit string $w\in\{0,1\}^m$ is divided into branches $w_0$ and $w_1$, 
with a corresponding rotation angle $\theta_{w,m}$. 
All rotations for strings of the same length $m$ can be applied in parallel; hence the main resource cost arises from constructing the angle-preparation oracle $U_m$.
\begin{Theorem}[Angle-preparation oracle $U_m$]\label{Th:U}
The unitary operator $U_m$ implements
\begin{equation}
\textstyle\sum\limits_{w\in \{0,1\}^m}\alpha_w|w\rangle|0\rangle\mapsto\textstyle\sum\limits_{w\in \{0,1\}^m}\alpha_w|w\rangle|\theta_{w,m}\rangle,
\end{equation} where the rotation angle $\theta_{w,m}$ is defined by
\begin{equation}
      \theta_{w,m}=\arcsin{\sqrt{S(a_1,b_1,\beta)/S(a_2,b_1,\beta)}},
\end{equation}
in which $S(a,b,\sigma)=\textstyle\sum_{n=a}^b n^{-\beta}$ and
\begin{align}
    a_1&=2^{k-m-1}(2w+1)+n_0,\notag\\
    b_1&=2^{k-m-1}(2w+2)-1+n_0,\notag\\
    a_2&=2^{k-m}w+n_0.
\end{align}
Then $U_m$ can be implemented such that each rotation angle is estimated to within error $\varepsilon/k$, using
\begin{equation}
\mathrm{Poly}(\log \varepsilon^{-1},\log n_1,\log|1-\beta|^{-1},\,\beta,\,v)
\end{equation}
gates and ancilla qubits, where $v$ denotes the number of significant digits used to represent $\beta$.
\end{Theorem}
\noindent
\emph{Proof sketch.}  
Each angle $\theta_{w,m}$ is obtained by a bit-wise bisection procedure.
At iteration $i$, a trial angle $\theta_{g,i}$ is represented with $i$ bits, and 
$\sin^2(\theta_{g,i})$ is  compared  to the target ratio $S(a_1,b_1,\beta)/S(a_2,b_1,\beta)$. This comparison determines the $i$-th decimal digit of $\theta_{w,m}$. Repeating this procedure for $\mathcal{O}(\log(k/\varepsilon))$ iterations successively refines the approximation, yielding $\theta_{w,m}$ with precision $\varepsilon/k$. Furthermore, we show that both the computation of $\sin^2(\theta_{g,i})$ and the partial sums  can be carried out with polynomial resources, which completes the proof.

\begin{Theorem}[Initial state preparation]\label{th:init}
Let $\beta>0,\;\beta \neq 1$ and $N\in\mathbb{N}$. Define
    \begin{equation}
        |\psi_0\rangle=\frac{1}{C}\textstyle\sum\limits_{n=1}^{N}n^{-\beta/2}|n\rangle, \qquad C=\sqrt{\textstyle\sum\limits_{n=1}^{N}n^{-\beta}}.
    \end{equation}
    Then  $|\psi_0\rangle$ can be prepared on a quantum computer to precision $\varepsilon>0$, with success probability at least $(\frac{1}{2}-\frac{\varepsilon}{3})$. The required number of gates and ancilla qubits is bounded by
    \begin{equation}
        \mathrm{Poly}( \log \varepsilon^{-1},\log N,\log|1-\beta|^{-1},\beta,v),
    \end{equation}
where $v$ denotes the number of significant digits used to represent $\beta$.
\end{Theorem}
\emph{Proof sketch.} The construction proceeds in three steps.  
First, prepare the auxiliary state using  controlled rotations~\cite{long2001efficient}, noted that $n_0$ has only linear reliance on $\log(\varepsilon^{-1})$ and $\beta$:
\begin{equation}
    {|\psi_{0}\rangle}_2=\frac{1}{C_2}\textstyle\sum\limits_{n=1}^{n_0-1}n^{-\beta/2}|n\rangle.
\end{equation}
Second, construct ${|\psi_{0}\rangle}_1$ using Theorem~\ref{th:tru},
and combine it with ${|\psi_{0}\rangle}_2$ via the LCU method, yielding
\begin{equation}
    {|\psi_{0}\rangle}_1+{|\psi_{0}\rangle}_2.
\end{equation}
Third, post-select to eliminate the states with index $n> N$, given by:
\begin{equation}
    {|\psi_{0}\rangle}_3=\frac{1}{C_3}\textstyle\sum\limits_{n=N+1}^{n_1-1}n^{-\beta/2}|n\rangle,
\end{equation}
yielding the desired state
\begin{equation}
    |\psi_{0}\rangle={|\psi_{0}\rangle}_1+{|\psi_{0}\rangle}_2-{|\psi_{0}\rangle}_3.
\end{equation}
By appropriately choosing $n_0$ and $n_1$, the  procedure succeeds with probability at least $(\frac{1}{2}-\tfrac{\varepsilon}{3})$ and requires only polynomial resources.

\subsubsection{Hamiltonian evolution}
\begin{Theorem}[Evolution operator construction]\label{Theorem:H}
    Define the time evolution operator
    \begin{equation}
        U(t)=\mathrm{e}^{-i\mathcal{H}_0t}, \qquad \mathcal{H}_0=\sum_{n=1}^{N}\log n|n\rangle\langle n|,
    \end{equation}
    where the evolution time $t$ is specified with $u$ significant digits. Then $U(t)$ can be implemented to precision $\xi$ using 
    \begin{equation}
        \mathrm{Poly}(\log N,\log |t|,\log\xi^{-1},u)
    \end{equation}
 gates and  ancilla qubits.
\end{Theorem}
\emph{Proof sketch.}  The construction begins by using the logarithm oracle from Lemma~\ref{le:log}, which maps 
\begin{equation}
    \textstyle\sum\limits_{n=1}^{N} \alpha_n|n\rangle|0\rangle \mapsto \textstyle\sum\limits_{n=1}^{N}\alpha_n|n\rangle|\widetilde{\log n}\rangle, 
\end{equation}
where $\widetilde{\log n}$ approximates $\log(n)$ to accuracy $O(\xi/|t|)$. The ancilla register storing $\widetilde{\log n}$ requires $\log N$ qubits for the integer part and $\log(|t|/\xi)$ qubits for the fractional part.  Controlled $R_z$ rotations conditioned on these qubits then imprint the phase factor $\mathrm{e}^{-it\log n}$ onto each computational basis state $|n\rangle$, thereby realizing $U(t)$ to precision $\xi$ with polynomial overhead in the stated parameters (see Supplementary Note 5 for details).\par
The implementation of  the controlled evolution
\begin{equation}
\exp(-i\mathcal{H}_{c2}t)
=\ket{\downarrow}\bra{\downarrow} \otimes U(t)
\end{equation}
in the first system, as well as the time evolution 
\begin{align}
       &\exp(-i\textstyle\int\mathcal{H}_{c}dt)\notag\\
     =&H(\mathrm{e}^{i\theta t}\ket{\uparrow}\bra{\uparrow} \otimes U(t)+\mathrm{e}^{-i\theta t}\ket{\downarrow}\bra{\downarrow} \otimes U^\dagger(t))H  
\end{align}
in the second system, exhibit the same complexity scaling. 
According to Theorem~\ref{Theorem:H}, both evolutions can be implemented to precision $\xi$ using
\begin{equation}
 \mathrm{Poly}\left(\log N, \log |t|, \log \xi^{-1}, u\right) 
\end{equation}
gates and  ancilla qubits.

\subsection{The computational complexity of calculating the  zeta Function in the critical strip}\label{Methods:complexity}
In classical computation, evaluating $\zeta(s)$ at  large imaginary parts in the critical strip $0<\beta<1$ is notoriously challenging.  
The Riemann-Siegel formula can be expressed as: 
\[
    \zeta(s)=\textstyle\sum\limits_{n=1}^{x}\frac{1}{n^s}+\chi(s)\textstyle\sum\limits_{n=1}^{y}\frac{1}{n^{1-s}}+O(x^{-\beta})+O(|t|^{1/2-\beta}y^{\beta-1}),
\]
with $2\pi xy=|t|$ and $\chi(s)=2^s\pi^{s-1}\sin(\tfrac{\pi s}{2})\Gamma(1-s)$. Setting $x=y=\sqrt{|t|/2\pi}$ yields
\begin{equation}
    \zeta(s)=\textstyle\sum\limits_{n=1}^{N}\limits n^{-s}+\chi(s)\textstyle\sum\limits_{n=1}^{N}n^{-(1-s)}+O(|t|^{-\beta/2}),
\end{equation}
where $N=\lceil\sqrt{|t|/2\pi}\rceil$. Thus, the main computational challenge reduces to evaluating partial Dirichlet sums.\par
Using the quantum constructions in previous sections, we prepare the initial state 
\begin{equation}
        |\psi_0\rangle=\frac{1}{C}\textstyle\sum\limits_{n=1}^{N}n^{-\beta/2}|n\rangle, \qquad C=\sqrt{\textstyle\sum\limits_{n=1}^{N}n^{-\beta}},
    \end{equation} 
  and apply the evolution operator 
  \begin{equation}
        U(t)=\mathrm{e}^{-i\mathcal{H}_0t}, \qquad \mathcal{H}_0=\sum\limits_{n=1}^{N}\log n|n\rangle\langle n|.
    \end{equation}
    The average accumulated phase factor can be obtained from
    the expectation values of $\sigma_x + i\sigma_y$ on an introduced probe qubit, yielding
    \begin{equation}
    {\cal L}(\beta,t)=\langle \psi_0|\mathrm{e}^{-i\mathcal{H}_0 t}|\psi_0\rangle=\frac{1}{C^{2}}\textstyle\sum\limits_{n=1}^{N}n^{-s}.
\end{equation}
For $0<\beta<1$, we have
\begin{equation}
C^2=\textstyle\sum\limits_{n=1}^{N} n^{-\beta}
\le 1 + \int_{1}^{N} x^{-\beta}\,dx<\frac{N^{1-\beta}}{1-\beta},
\end{equation} 
and $|\chi(s)|=\Theta(|t|^{1/2 - \beta})$ (see Supplementary Note 7).
Hence, to approximate $\zeta(\beta+it)$ with precision $\delta$, it suffices to estimate the two partial sums $\textstyle\sum_{n=1}^{N}n^{-s}$ and $\textstyle\sum_{n=1}^{N}n^{-(1-s)}$ to precisions $\mathcal{O}(\delta)$  and $\mathcal{O}(\delta)|t|^{\beta-1/2}$ respectively. Therefore, the required precisions in the estimation of  $\langle\sigma_x\rangle$  and $\langle \sigma_y \rangle$ of the probe qubits are $d_1=(1-\beta)\mathcal{O}(\delta)/N^{1-\beta}$ and $d_2=\beta\mathcal{O}(\delta)|t|^{\beta-\tfrac{1}{2}}/N^{\beta}$, respectively. 
To achieve an overall precision of $d_1$ and $d_2$, the preparation of the initial state, the time evolution, and sampling the outputs must each be implemented to accuracies of order $\mathcal{O}(d_1)$ and $\mathcal{O}(d_2)$. By Theorem \ref{th:init} and Theorem \ref{Theorem:H}, substituting $\varepsilon$ and $\xi$ with $\mathcal{O}(d_1)$ and $\mathcal{O}(d_2)$ yield gate and ancilla resource requirements for initial state preparation and evolution  bounded by 
\begin{equation}
         \mathrm{Poly}(\log\delta^{-1},\log|t|,\log(1-\beta)^{-1},\log\beta^{-1},v),
\end{equation}
and 
\begin{equation}
\mathrm{Poly}(\log\delta^{-1},\log|t|,\log(1-\beta)^{-1},\log\beta^{-1},u),
\end{equation}
respectively, where $u,v$ denote the number of significant digits of $t$ and $\beta$. \par
Let  $\zeta'(s)$  denote the derivative of $\zeta(s)$, which can be bounded by
    $\mathrm{Poly}(|t|,|1-\beta|^{-1},\beta^{-1})$ (see Supplementary Note 6 for details). Thus, setting $u,v$  as $\mathrm{Poly}(\log\delta^{-1},\log|t|,\log(1-\beta)^{-1},\log\beta^{-1})$
ensures that rounding errors contribute at most $\mathcal{O}(\delta)$. Substituting these bounds into the previous bounds gives gate  and ancilla qubit costs $\mathcal{R}_{c1}$ and $\mathcal{R}_{c2}$ for both summations:
\begin{equation}\label{eq:initial complexity2}
\mathrm{Poly}(\log\delta^{-1},\log|t|,\log(1-\beta)^{-1},\log\beta^{-1}).
\end{equation}\par
If  $\langle\sigma_x\rangle$  and $\langle \sigma_y \rangle$ are estimated by direct sampling, sample complexity scales as $\mathcal{O}(d_1^{-2})$ and $\mathcal{O}(d_2^{-2})$. With amplitude amplification or quantum amplitude estimation, this can be reduced to $\mathcal{O}(d_1^{-1})$ and $\mathcal{O}(d_2^{-1})$. Since $N=\Theta(\sqrt{|t|})$, the required number of samples is bounded by
\begin{align}
    \mathcal{R}_{s1}&=\mathcal{O}(\delta^{-1})(1-\beta)^{-1}|t|^{(1-\beta)/2},\notag\\
    \mathcal{R}_{s2}&=\mathcal{O}(\delta^{-1})\beta^{-1}|t|^{(1-\beta)/2}.
\end{align}
Hence, the overall complexity
$\mathcal{R}_{o}=\mathcal{R}_{c1}\mathcal{R}_{s1}+\mathcal{R}_{c2}\mathcal{R}_{s2}$ is bounded by:
\begin{equation}
    \delta^{-1}|t|^{(1-\beta)/2}
    \mathrm{Poly}(\log\delta^{-1},\log|t|,(1-\beta)^{-1},\beta^{-1}).
\end{equation}
\par

Next, we consider the task of locating and verifying nontrivial zeros. These zeros are symmetric with respect to the critical line $\beta = \tfrac{1}{2}$, and there is a known zero-free region, known as Korobov--Vinogradov zero-free region, which takes the form~\cite{mossinghoff2024explicit}:
\begin{equation}
    \beta > 1 - \frac{1}{ \mathcal{O}((\log |t|)^{2/3} (\log \log |t|)^{1/3})}.
\end{equation}
 Thus, we focus on the region satisfying
\begin{equation}\label{eq:region}
    \tfrac{1}{2}\leq \beta \leq 1-\frac{1}{\mathcal{O} (\, (\log |t|)^{2/3} (\log \log |t|)^{1/3})},
\end{equation}
for which the complexity bound simplifies to
\begin{equation}
    \delta^{-1}|t|^{(1-\beta)/2}
    \mathrm{Poly}(\log\delta^{-1},\log|t|).
\end{equation}


\section*{Data and code availability} 
Data and code are available from the corresponding author upon reasonable request.
\section*{Acknowledgements} 

S.W. acknowledges  Beijing Nova Program (Grants No. 20230484345). This work is supported by the National Natural Science Foundation of China (62571050, 12275117), Guangdong Basic and Applied Basic Research Foundation (2022B1515020074), and Shenzhen Science and Technology Program (RCYX20200714114522109 and KQTD20200820113010023). We thank Feihao Zhang for the helpful discussion. 

\section*{Author contributions}
T.X.,  G.L.L. and S.W. supervised the project. S.W. proposed the initial theoretical idea and developed the theoretical framework together with Q.L. and T.X.. Y.Z. performed the experiments under the supervision of T.X. W.Y., P.G., F.N. contributed to the theoretical analysis, while C.W. and J.S. contributed to the experimental work. All authors participated in the preparation of the manuscript.
\section*{Competing interests}
The authors declare no competing interests.
\newpage
\begin{figure*}[hbt!]
    \begin{center}
    \includegraphics[width=0.995\linewidth]{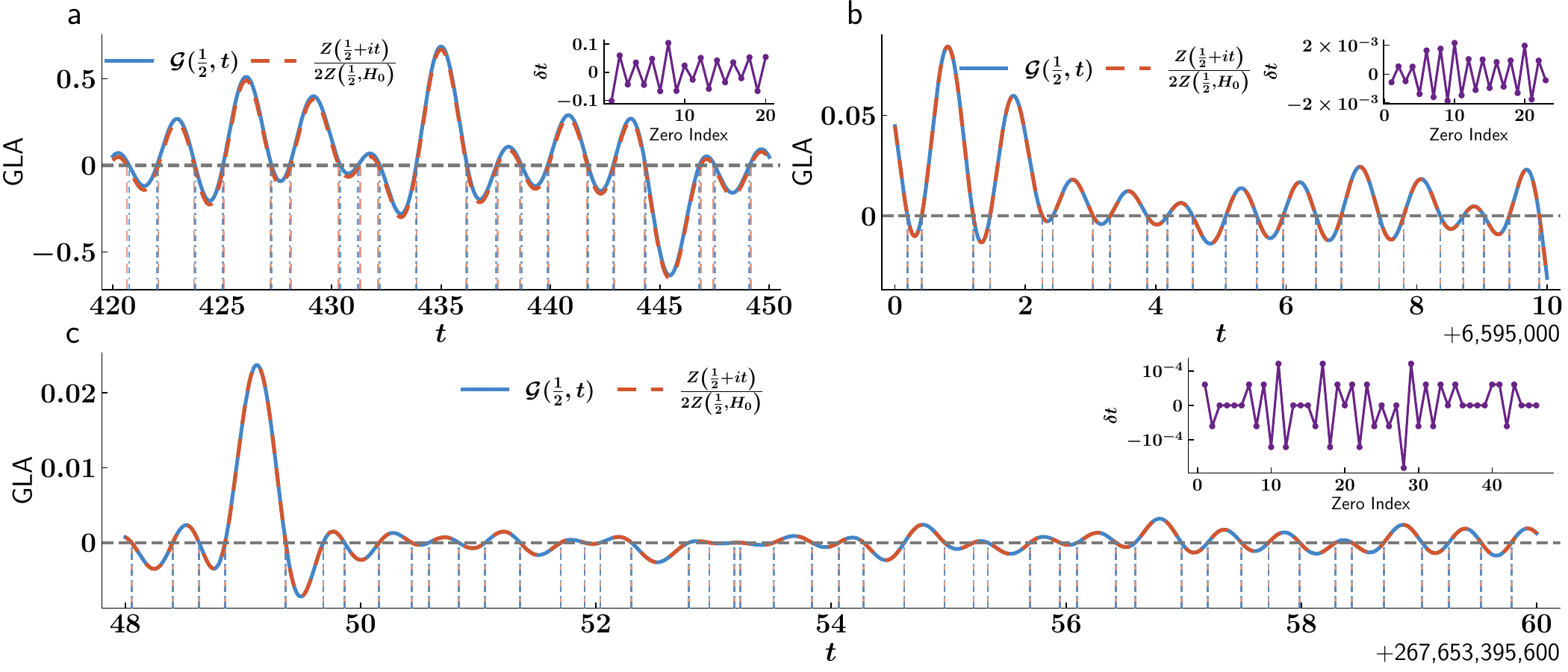}
    \end{center}
    \vspace{-0.50cm}
    \caption{\textbf{Generalized Loschmidt amplitude ${\cal G}(\tfrac{1}{2},t)$ across different imaginary part regions.} \textbf{a} Nontrivial  zeros with imaginary parts in the range $[420,450]$, corresponding to the $216$-th to $235$-th zeros, obtained using a a 3-spin quantum system.
\textbf{b} Nontrivial zeros with imaginary parts in the range $[6.595\times10^{6},6.595\times10^{6}+10]$, corresponding to the $\num{13502344} $-th to $\num{13502366}$-th zeros, obtained using a 10-spin quantum system. \textbf{c} Nontrivial zeros with imaginary parts in the range $[\num{267653395648},\,\num{267653395660}]$, corresponding to the $(10^{12}-3)$-th to $(10^{12}+43)$-th  zeros, obtained using a 18-spin quantum system. The blue solid line shows the simulated ${\cal G}(\tfrac{1}{2},t)$, while the red dashed line show $Z(\tfrac{1}{2}+it)/\left(2\mathcal{Z}(\tfrac{1}{2},H_0)\right)$.
Vertical red dashed lines mark the exact Riemann zeros, and blue dashed and dotted lines mark the zeros estimated by the quantum system. Insets show the deviation $\delta t$ between the exact zeros and estimated zeros, which  decreases with increasing $t$.} 
    \vspace{-0.50cm}
    \label{EFig1}
\end{figure*}

\begin{table*}[htbp!]
\centering
\renewcommand{\arraystretch}{1.9}
\begin{tabular}{|c|c|c|}
\hline
\textbf{Region of $\beta$} & \textbf{Sample complexity $\mathcal{R}_s$} & \textbf{Total complexity} $\mathcal{R}_c\mathcal{R}_s$ \\
\hline
$0<\beta<1$ & 
$\mathcal{O}(\delta^{-1})|t|^{(1-\beta)/2}(\beta^{-1}+(1-\beta)^{-1})$& 
$    \delta^{-1}|t|^{(1-\beta)/2}
    \mathrm{Poly}(\log \delta^{-1},\log |t|,(1-\beta)^{-1},\beta^{-1})$ \\
\hline
Possible zero region& 
$\mathcal{O}(\delta^{-1})|t|^{(1-\beta)/2}(\beta^{-1}+(1-\beta)^{-1})$ & 
$    \delta^{-1}|t|^{(1-\beta)/2}
    \mathrm{Poly}(\log \delta^{-1},\log |t|)$ \\
\hline
$\beta=\tfrac{1}{2}$& 
$\mathcal{O}(\delta^{-1})|t|^{1/4}$ & 
$    \delta^{-1}|t|^{1/4}
    \mathrm{Poly}(\log \delta^{-1},\log( |t|)$ \\
\hline
\textbf{Classical method} & \multicolumn{2}{c|}{$\sim|t|$ (Alternating series)~\cite{borwein2000efficient}, $\sim|t|^{1/2}$ (Riemann-Siegel formula)~\cite{de2011high}}\\
\hline
\end{tabular}
\caption{\textbf{Complexity bounds of the quantum algorithm for evaluating $\zeta(\beta+it)$ with precision $\delta$ across different regions, compared with classical approaches.} The ``possible zero region'', defined in Eq.~\eqref{eq:region}, is the  regime of interest for locating and verifying nontrivial zeros.}
\label{tab:complexity}
\end{table*}
\begin{figure*}[hbt!]
    \begin{center}
    \includegraphics[width=0.98\linewidth]{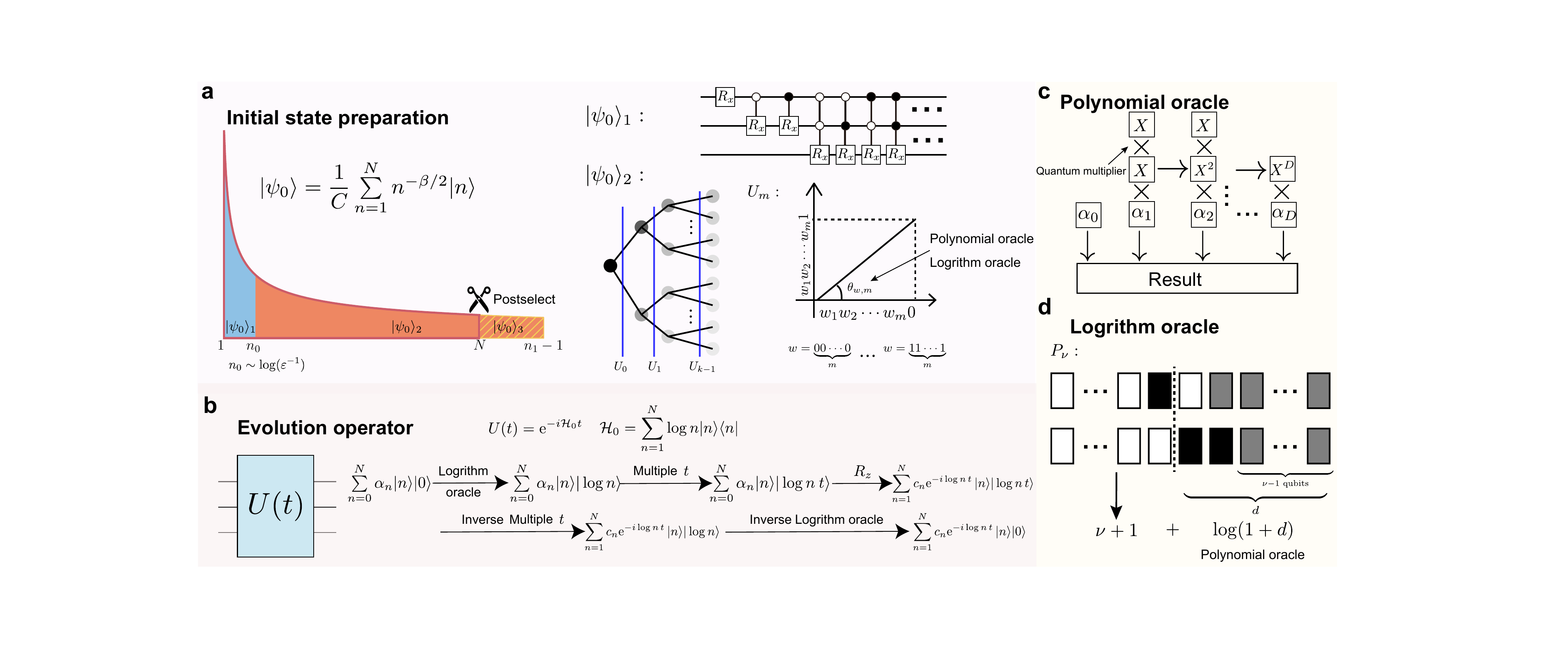}
    \end{center}
    \vspace{-0.50cm}
\caption{\textbf{Construction of the quantum system on a gate-based quantum computer.} 
\textbf{a}, The initial state is prepared as $|\psi_0\rangle = |\psi_0\rangle_1 + |\psi_0\rangle_2 - |\psi_0\rangle_3$ (normalization omitted). 
The components $|\psi_0\rangle_1$, $|\psi_0\rangle_2$, and $|\psi_0\rangle_3$ correspond to the blue, orange, and slashed yellow regions, respectively. 
The states $|\psi_0\rangle_1$ and $|\psi_0\rangle_2$ are generated using controlled rotations and Grover’s amplitude-splitting method, respectively. 
\textbf{b}, The time-evolution operator is realized in five sequential steps. 
\textbf{c}, \textbf{d}, The polynomial and logarithmic oracles serve as preliminary blocks for implementing the time-evolution operator and preparing the initial state.}
    \vspace{-0.50cm}
    \label{EFig2}
\end{figure*}
\clearpage
\onecolumngrid
\setcounter{section}{0}
\setcitestyle{super,sort&compress}
\renewcommand{\thesection}{Supplementary Note \arabic{section}}

\numberwithin{equation}{section}
\renewcommand{\theequation}{\arabic{section}.\arabic{equation}}
\makeatletter 
\makeatother
\newtheorem{theorem}{Theorem}[section]
\newtheorem{lemma}[theorem]{Lemma}
\newtheorem{proposition}[theorem]{Proposition}
\newtheorem{corollary}[theorem]{Corollary}
\newtheorem{definition}[theorem]{Definition}
\renewcommand{\thetheorem}{\arabic{section}.\arabic{theorem}} 
\renewcommand{\thepage}{S\arabic{page}} 

\setcounter{page}{1}
\renewcommand{\thepage}{S\arabic{page}} 
\begin{center}
   \large \bfseries Supplementary information for "The Riemann Hypothesis Emerges in Dynamical Quantum Phase Transitions"
\end{center}

\titleformat{\section}
{\centering\normalsize\bfseries}
{\thesection:}
{0pt}
{\normalsize}
\begin{center}\textbf{Notation}
\end{center}\par
For clarity, throughout this Supplementary Information, we use `$\log$' to denote the base-2 logarithm, and `$\ln$' to denote the natural logarithm (base $e$). All variables $s$ are assumed to satisfy $\Re(s) > 0$ and $\Re(s)\neq1$, ensuring the applicability of the alternating Dirichlet series. $\beta$ denote the real part of $s$ and $t$ denote the imaginary part of $s$.\par
The Big-O notation $\mathcal{O}(\cdot)$ represents the worst-case upper bound, while the Big-Theta notation $\Theta(\cdot)$ denotes the exact asymptotic behavior. The notation $\text{Poly}(\cdot)$ signifies that the growth follows a polynomial function of the input.\par
The symbol $\|\cdot\|$ denotes the Euclidean norm on state vectors. 
The distance between two pure quantum states $|\psi_{a}\rangle$ and $|\psi_{b}\rangle$ is defined as $d(|\psi_a\rangle,|\psi_b\rangle)
= \displaystyle \min\limits_{\theta \in \mathbb{R}}
\||\psi_a\rangle - e^{i\theta}|\psi_b\rangle\|
= \sqrt{2(1-|\langle\psi_a|\psi_b\rangle|)}$, which satisfies the triangle inequality. The distance between two unitary operators is defined as $D(U_1,U_2)=
\displaystyle\max\limits_{|\psi\rangle}d(U_1|\psi\rangle,U_2|\psi\rangle)$, which is the worst-case distance over all input states $|\psi\rangle$. We say that a quantum state (or unitary operator) is prepared to precision $\kappa$ if its distance from the corresponding ideal target state (or ideal unitary) is at most $\kappa$.\par 
\section{Asymptotic behavior of $S_{N}(s)$}
The $N$-term partial sum of the alternating Dirichlet series is 
\begin{equation}
    S_N(s)=\sum_{n=1}^{N}\frac{(-1)^{n-1}}{n^s}.
\end{equation}\par
\textbf{Case~1.}  At the nontrivial zeros of the  zeta function, we have $S_N(s) = -R_N(s)$, where
\begin{equation}
    R_N(s)=\sum_{n=N+1}^{\infty}\frac{(-1)^{n-1}}{n^s}.
\end{equation}
Each term has the following integral representation,
\begin{equation}
   \frac{1}{n^s} = \frac{1}{\Gamma(s)} \int_0^{\infty} \tau^{s-1} e^{-n\tau} d\tau,
\end{equation}
yielding
\begin{equation}
    R_N(s) = \sum_{n=N+1}^{\infty} \frac{(-1)^{n-1}}{n^s} = \frac{1}{\Gamma(s)} \int_0^{\infty} \tau^{s-1} \sum_{n=N+1}^{\infty} (-1)^{n-1} e^{-n\tau} d\tau=\frac{(-1)^N}{\Gamma(s)} \int_0^{\infty} \frac{\tau^{s-1} e^{-(N+1)\tau}}{1+e^{-\tau}} d\tau.
\end{equation} 
Using the identity
\begin{equation}
    \frac{1}{1+e^{-\tau}} = \frac{1}{2} + \frac{1}{2} \tanh\left(\frac{\tau}{2}\right),
\end{equation}
we obtain
\begin{equation}\label{eq:MJ}
    R_N(s) = \frac{(-1)^N}{\Gamma(s)} \Bigl[\frac{1}{2}\int_0^{\infty}\tau^{s-1} e^{-(N+1)\tau}d\tau+\frac{1}{2}\int_0^{\infty}\tau^{s-1} e^{-(N+1)\tau}\tanh( \frac{\tau}{2})d\tau\Bigr].
\end{equation}
Let us define
\begin{equation}
    M:=\frac{1}{2}\int_0^\infty \tau^{s-1}e^{-(N+1)\tau}d\tau=\frac{1}{2}\Gamma(s)(N+1)^{-s}, \qquad J:=\frac{1}{2}\int_0^\infty \tau^{s-1}e^{-(N+1)\tau}\tanh( \frac{\tau}{2})d\tau.
\end{equation}
This leads to
\begin{equation}\label{eq:sum}
R_N(s)=\frac{(-1)^N}{\Gamma(s)}\big(M+J\big)=(-1)^N\Big(\frac{1}{2(N+1)^s}+\frac{J}{\Gamma(s)}\Big),
\end{equation}
We now show that for sufficiently large $N$, 
\begin{equation}
    |J|\leq\frac{1}{4}\Gamma(s)(N+1)^{-\beta}, \qquad s=\beta+it.
\end{equation}\par
 Fix $A>0$, which depends on $s$ but not on $N$, and split the integral for $J$ at $\tau = A/(N+1)$:
\begin{equation}
    J=\frac{1}{2}\Bigg(\int_0^{A/(N+1)}\tau^{s-1}e^{-(N+1)\tau}\tanh( \frac{\tau}{2})d\tau+\int_{A/(N+1)}^\infty \tau^{s-1}e^{-(N+1)\tau}\tanh( \frac{\tau}{2})d\tau\Bigg) 
:= \frac12(I_{1}+I_{2}).
\end{equation}
For $I_1$, we apply the inequality \( |\tanh(\tau/2)| \le |\tau/2| \), so
\begin{equation}
|I_1|\le\int_0^{A/(N+1)} \tau^{\beta-1} e^{-(N+1)\tau}\frac{\tau}{2}d\tau
=\frac{1}{2} (N+1)^{-(\beta+1)}\int_0^A u^{\beta} e^{-u}du=C_1(\beta,A)(N+1)^{-(\beta+1)}.
\end{equation}
Here,  $C_1(\beta,A):=\tfrac12\int_0^A u^{\beta}e^{-u}du$ is a constant  depending on $A$ and $\beta$. For $I_2$, we use the trivial bound $|\tanh(\tau/2)|\le 1$, so
\begin{equation}
    |I_2|\le \int_{A/(N+1)}^\infty \tau^{\beta-1} e^{-(N+1)\tau}d\tau=(N+1)^{-\beta}\int_A^\infty u^{\beta-1} e^{-u}du=C_2(\beta,A)(N+1)^{-\beta}.
\end{equation}
Here $C_2(\beta,A):=\int_A^\infty u^{\beta-1}e^{-u}du$ is also a constant that depends on $A$ and $\beta$.\par
Since $\Gamma(s)$ is non-vanishing  and $C_2(\beta,A)\to 0$ as $A\to\infty$, we can choose $A$ large enough so that
\begin{equation}
    C_2(\beta,A)\le \frac{1}{4}|\Gamma(s)|.
\end{equation}
Thus,
\begin{equation}
    |I_{2}|\leq\frac{1}{4}|\Gamma(s)|(N+1)^{-\beta}.
\end{equation}
Combining the two contributions gives
\begin{equation}
    |J|\le \frac12\big(|I_{1}|+|I_{2}|\big)
\le \frac12\Big(C_1(\beta,A)(N+1)^{-(\beta+1)}+\frac{1}{4}|\Gamma(s)|(N+1)^{-\beta}\Big).
\end{equation}
Consequently, for all $N\ge N_0:=\bigl\lceil 4C_1(\beta,A)/|\Gamma(s)|\bigr\rceil$, 
\begin{equation}
    |J|\le \frac{1}{4}|\Gamma(s)|(N+1)^{-\beta}, \qquad \Bigl|\frac{J}{\Gamma(s)}\Bigr|\leq \frac{1}{4}(N+1)^{-\beta}.
\end{equation}
Substituting this bound into Equation \eqref{eq:sum} gives for $N\geq N_0$,
\begin{equation}
    \frac{1}{4}(N+1)^{-\beta}\leq|R_N(s)|\leq\frac{3}{4}(N+1)^{-\beta}.
\end{equation}
Therefore,
\begin{equation}
    \lim_{N \to \infty}
    -\frac{1}{\log N}\ln |S_N(s)| =\lim_{N \to \infty}
    -\frac{1}{\log N}\ln |R_N(s)| = \beta\ln2.
\end{equation}\par
\textbf{Case~2.}  For $s$ with $\zeta(s)\neq 0$, the partial sums converge, and in particular
\begin{equation}
    \lim_{N\to\infty} S_N(s)
    = (1-2^{\,1-s})\,\zeta(s)
    \neq 0.
\end{equation}
It follows that
\begin{equation}
    -\lim_{N\to\infty}\frac{1}{\log N}\,\ln|S_N(s)|=0.
\end{equation}
Thus, we obtain the final result:
\begin{equation}
-\lim_{N\to\infty}\frac{1}{\log N}\,\ln |S_N(s)|
=\begin{cases}
\beta\ln 2, & \text{if $\zeta(s)=0$},\\[2pt]
0, & \text{if $\zeta(s)\neq 0$}.
\end{cases}
\end{equation}
\section{Asymptotic behavior of $\mathcal{Z}(\beta,\mathcal{H}_0)$}
The partition function can be written as
\begin{equation}
    \mathcal{Z}(\beta,\mathcal{H}_0) = \sum_{n=1}^{N} n^{-\beta}.
\end{equation}\par
\textbf{Case~1.} For $\beta>1$, $\mathcal{Z}(\beta,\mathcal{H}_0)$ converges as
\begin{equation}
    \lim_{N \to \infty}
    \mathcal{Z}(\beta,\mathcal{H}_0)
    = \zeta(\beta).
\end{equation}
Thus,
\begin{equation}
    \lim_{N \to \infty}
    -\frac{1}{\log N}\ln \mathcal{Z}(\beta,\mathcal{H}_0)  = 0.
\end{equation}\par
\textbf{Case~2.} For $0 < \beta < 1$, the series diverges. Specifically,
\begin{equation}\label{eq:integral-bounds}
\int_{1}^{\,N+1} x^{-\beta}\,dx \;\le\; \sum_{n=1}^{N} n^{-\beta} \;\le\; 1+\int_{1}^{\,N} x^{-\beta}\,dx,
\end{equation}
which simplifies to
\begin{equation}\label{eq:asymp-sum}
\frac{(N+1)^{1-\beta}-1}{1-\beta}\;\le\;\mathcal Z(\beta,\mathcal{H}_0)\;\le\;1+\frac{N^{1-\beta}-1}{1-\beta},
\end{equation}
 yielding
\begin{equation}
    -\lim_{N\rightarrow\infty}\frac{1}{\log N}\ln\mathcal Z(\beta,\mathcal{H}_0)=(\beta-1)\ln2.
\end{equation}
Thus,
\begin{equation}
-\lim_{N\to\infty}\frac{1}{\log N}\,\ln \mathcal{Z}(\beta,\mathcal{H}_0)
=\begin{cases}
(\beta-1)\,\ln 2, & 0<\beta<1,\\[2pt]
0, & \beta> 1.
\end{cases}
\end{equation}

\section{Construction of the preliminary  oracles}
This section details the construction of polynomial and logarithm oracles, which serve as foundational components for subsequent quantum state preparation and time evolution operators in our quantum algorithms.
\begin{lemma}[Polynomial oracle (Methods, Lemma~1)]\label{polynomial}
Let $O(f)$ denote an  oracle implementing the transformation 
\begin{equation}
    \sum_x\alpha_x|x\rangle|0\rangle\mapsto \sum_x\alpha_x|x\rangle|f(x)\rangle,
\end{equation}
where $f(x)$ is a polynomial of degree at most $D$. Suppose the coefficients of $f$ and the input $x$ are specified to $a_1$ and $a_2$ significant digits, respectively. The output is encoded with $r_1$ integer qubits and $r_2$ fractional qubits. Then  $O(f)$ can be implemented using
\begin{center}
 $\mathcal{O}\!\left(D^{2}a_2^{2} + D^{2}a_1a_2 + D(r_1+r_2)\right)$ gates, and $(2Da_2+a_1)$ ancilla qubits.   
\end{center}
\end{lemma}
\noindent\textit{Proof.} 
Write the polynomial as $f(n)=\textstyle\sum_{d=0}^{D} c_d n^{d}$. Evaluation proceeds through sequential construction of the monomials $x^{d}$, multiplication by coefficients $c_d$, and accumulation of the results.

\textbf{Step~1: Computation of Monomials.}
Multiplying an $m_1$-qubit register by an $m_2$-qubit register requires $\mathcal{O}(m_1 m_2)$ gates and $\mathcal{O}(m_1 + m_2)$ qubits using quantum schoolbook multiplication~\cite{litinski2024quantum}. To compute \( x^{d+1} \) from \( x^d \), a $d a_2$-qubit register (storing \( x^d \)) is multiplied by an $a_2$-qubit register (storing \( x \)). The total gate cost for generating monomials up to degree \( D \) is
\begin{equation}
2 \sum_{d=1}^{D-1} a_2 \cdot d a_2 = \mathcal{O}(D^{2}a_2^{2}),
\end{equation}
where the factor $2$ arises from uncomputing intermediate results and resetting ancilla qubits. \par
\textbf{Step~2: Multiplication by coefficients.}  
 Multiplying a degree-\( d \) monomial (which requires \( d a_2 \) qubits) by \( c_d \) costs \( \mathcal{O}(d a_1 a_2) \) gates. 
Summing over all degrees gives
\begin{equation}
2 \sum_{d=1}^{D} d a_1a_2 = \mathcal{O}(D^2a_1a_2),
\end{equation}
again with the factor 2 from uncomputation.  

\textbf{Step~3: Accumulation of output.} 
The results are accumulated into an output register of size \( r_1 + r_2 \). Each of the \( D + 1 \) additions costs \( \mathcal{O}(r_1 + r_2) \) gates, yielding a total gate cost of \( \mathcal{O}(D (r_1 + r_2)) \). 
Adding all contributions yields a total gate complexity
\begin{equation}
\mathcal{O}\!\left(D^{2}a_2^{2} + D^{2}a_1a_2 + D(r_1+r_2)\right).
\end{equation}\par
During computation, at most  \( D a_2 \) qubits are used for monomial storage and 
\( (D a_2 + a_1) \)  for intermediate products, giving a total ancilla requirement of 
 \( (2D a_2 + a_1) \).\qed

 \begin{lemma}[Logarithm oracle (Methods, Lemma~2)]\label{the:log}
    Let $L$ denote an oracle implementing the transformation 
    \begin{equation}
        \sum_{n=1}^{N} \alpha_n|n\rangle|0\rangle \mapsto \sum_{n=1}^{N}\alpha_n|n\rangle|\widetilde{\log(n)}\rangle,
    \end{equation}
    where $\widetilde{\log(n)}$ approximates $\log(n)$ to within error $\eta$. Then $L$ can be implemented using 
       $\mathcal{O}((\log N)^3 \log^2(1/\eta))$
    gates and $\mathcal{O}((\log N)^2 \log(1/\eta))$ ancilla qubits. 
\end{lemma}
\noindent\textit{Proof.} We proceed constructively. 

\textbf{Step~1: Input Partitioning.}  
Define $k_1=\lceil\log(\tfrac{N+1}{3})\rceil$. For $3\leq n <3\cdot 2^{k_1}$, partition the integers into $k_1$ subsets, defined as
\begin{equation}
    P_\nu : \quad 3\cdot 2^{\nu-1} \leq n_\nu < 3\cdot 2^\nu, \qquad 1 \leq \nu \leq k_1.
\end{equation}
Each $3\leq n \leq N$ lies in exactly one partition $P_\nu$.
For $n_\nu \in P_\nu$, define
\begin{equation}
    d = \frac{n_\nu-2^{\nu+1}}{2^{\nu+1}}, \qquad -\frac{1}{4} \leq d < \frac{1}{2}.
\end{equation}
This allows the logarithm to be expressed as
\begin{equation}
    \log(n_\nu) = \log(2^{\nu+1}) + \log(1+d).
\end{equation}\par
\textbf{Step~2: Taylor expansion of \boldmath$\log(1+d)$.}  
Using the Taylor series expansion for \( \log(1 + d) \), we have:
\begin{equation}
    \log(1+d) = \frac{1}{\ln 2} \sum_{j=1}^\infty \frac{(-1)^{j+1}}{j} d^j.
\end{equation}
The remainder after $l_1$ terms satisfies
\begin{equation}
    \Biggl|\sum_{j=l_1+1}^{\infty} \frac{(-1)^{j+1}}{j} d^j \Biggr|
    \leq \sum_{j=l_1+1}^{\infty} \frac{1}{j} \Bigl(\frac{1}{2}\Bigr)^j
    \leq \frac{1}{2^{l_1}}.
\end{equation}
Hence, truncating after $l_1 =\lceil \log(1/\eta)\rceil+2$ terms yields an approximation error of less than \( \frac{\eta}{2} \). Coefficients stored to 
\( \lceil \log(2/\eta) \rceil \)  fractional bits introduce an additional rounding error bounded by
\begin{equation}
    \frac{\eta}{2}(1+\frac{1}{2}+(\frac{1}{2})^2+\cdots)\leq\frac{\eta}{2},
\end{equation}
giving total error of less than \( \eta \).
The resulting approximation
\begin{equation}\label{eq:taylor}
    \widetilde{\log(n_\nu)} = \nu+1 + \frac{1}{\ln(2)} \sum_{j=1}^{l_1} \frac{(-1)^{j+1}}{j} d^j
\end{equation}
is thus approximated within error $\eta$.

The partition index $\nu$ is determined using multi-controlled operations that compare $n$ with partition bounds. Each such comparison uses at most \( \mathcal{O}(k_1) \) Toffoli gates and \( \mathcal{O}(k_1) \) ancilla qubits. A flag register composing of at most $\mathcal{O}(k_1)$ qubits marks the unique partition $P_\nu$ containing $n$. This flag controls addition of the corresponding polynomial output to the output register. Exactly one flag is set for each input across $3\leq n \leq N$.\par
For each $\nu$, Equation~\eqref{eq:taylor} is a degree-\( l_1 \) polynomial in \( d \), with coefficients specified to \( \lceil \log(2/\eta) \rceil \) qubits and input \( d \) to \( (k_1 + 2) \) significant digits. The output requires \( (\lceil \log k_1 \rceil + 2) \) integer qubits and \( \lceil \log(1/\eta) \rceil \) fractional qubits. Applying Lemma~\ref{polynomial}, evaluating this polynomial within a partition requires
\begin{equation}
    \mathcal{O}(l_1^2 k_1^2 + l_1^2 k_1 \log(2/\eta) + l_1(\log(k_1)+\log(1/\eta))) = \mathcal{O}( k_1^2\log^2(1/\eta))
\end{equation}
gates, and
\begin{equation}
    2l_1k_1 + \log(1/\eta) = \mathcal{O}(k_1\log(1/\eta))
\end{equation}
ancilla qubits.
Summing over all $k_1$ partitions gives total resource requirements:
\begin{equation}
    \mathcal{O}(k_1^3 \log^2(1/\eta))=\mathcal{O}((\log N)^3 \log^2(1/\eta)) \;\text{gates}, \qquad
    \mathcal{O}(k_1^2 \log(1/\eta))=\mathcal{O}((\log N)^2 \log(1/\eta)) \; \text{ancilla qubits}.
\end{equation}
The special cases $n=1,2$ can be handled separately at negligible cost.\qed

\section{Construction of the initial state}
This section details the construction of the initial state, which is outlined as follows:\par
Our first key result is provided in  Theorem \ref{th:trusi} (Methods, Theorem 1), which describes the preparation  of the truncated state $|\psi_0\rangle_1$. Central to this construction is the angle-preparation oracle (Methods, Theorem 2) that computes the required rotation angles. Its action is defined in Definition \ref{def:Um} and its implementation  is outlined in Theorem \ref{the:Um}. A central subroutine of this oracle is the evaluation of a zeta function related partial sum (Lemma \ref{S oracle}), built upon Proposition \ref{lm:split} with detailed calculations given in Propositions \ref{prop:endpoint} and \ref{prop:summation}.\par
Our second key result, Theorem \ref{th:initsi} (Methods, Theorem 3),  presents the construction of our initial state $|\psi_0\rangle$. Starting with the truncated state $|\psi_0\rangle_1$ prepared in Theorem \ref{th:trusi}, we extend it via the Linear Combination of Unitaries (LCU) method to combine it with $|\psi_0\rangle_2$ (Corollary \ref{co:extended}); a final post-selection removes the high-index component $|\psi_0\rangle_3$ , yielding the desired initial state.
\begin{definition}
    The controlled-rotation gate $\mathcal{CR}$ implements the transformation
    \begin{equation}
\sum_{\theta}\alpha_{\theta}|\theta\rangle|0\rangle \;\mapsto\; \sum_{\theta}\alpha_{\theta}|\theta\rangle\bigl(\cos\theta\,|0\rangle + \sin\theta\,|1\rangle\bigr).
    \end{equation}
It can be implemented using a sequence of controlled-\(R_x\) rotations, each conditioned on a qubit of the register encoding \(\theta\), followed by a phase gate on the target qubit.
\end{definition}
\begin{theorem}[Truncated state preparation (Methods, Theorem~1)]\label{th:trusi}
    Let $\beta>0,\;\beta \neq 1$, and let $n_0,n_1\in\mathbb{N}$ with $n_1-n_0=2^k$ for some integer $k>0$. Define
    \begin{equation}
        |\psi_0 \rangle_1 \;=\; \frac{1}{C_1}\sum_{n=n_0}^{n_1-1} n^{-\beta/2}\,|n-n_0\rangle,
    \qquad 
        C_1=\sqrt{ \sum_{n=n_0}^{n_1-1} n^{-\beta}}.
    \end{equation}
    Then  $|\psi_0\rangle_1$ can be prepared on a standard gate-based quantum computer to precision $\varepsilon>0$, using a number of gates and ancilla qubits bounded by 
    \begin{equation}
        \mathrm{Poly}\!\left(\log(1/\varepsilon), \; \log(n_1), \; \log(\tfrac{1}{|1-\beta|}), \; \beta, \; v \right),
    \end{equation}
    where $v$ denotes the number of significant digits used to represent $\beta$. The parameters are required to satisfy
    \begin{equation}
        c = \Biggl\lceil \frac{1}{2}\log_{2\pi}\!\Bigl(\frac{8}{\varepsilon}\Bigr)\Biggr\rceil, 
        \qquad n_0 > \lceil \beta + 2c \rceil.
    \end{equation}
\end{theorem}
\noindent\textit{Proof sketch.}
We adapt Grover’s recursive amplitude-splitting method~\cite{grover2002creating}, replacing integration with partial summation.  The $2^k$ computational basis states are encoded on $k$ qubits initialized in $|0\rangle^{\otimes k}$. At iteration step $m$ ($0\le m\le k-1$), each bit string $w\in\{0,1\}^m$ is split into two branches $w0$ and $w1$. Define
\begin{equation}
    c_w=\frac{1}{C}\sqrt{\sum_{i=0}^{2^{k-m}-1}(2^{k-m}w+i+n_0)^{-\beta}}.
\end{equation}
Normalization is preserved, as
\begin{equation}
    c_{w0}=\frac{1}{C}\sqrt{\sum_{i=0}^{2^{k-m-1}-1}\!\bigl(2^{k-m}w+i+n_0\bigr)^{-\beta}}, 
    \quad
    c_{w1}=\frac{1}{C}\sqrt{\sum_{i=0}^{2^{k-m-1}-1}\!\bigl(2^{k-m}(w+1)+i+n_0\bigr)^{-\beta}},
\end{equation}
satisfy $c_{w0}^2+c_{w1}^2=c_w^2$. For each $w$, we compute a rotation angle $\theta_{w,m}$ such that the amplitude split $c_w\mapsto (c_{w_0},c_{w_1})$ can be implemented by a $\mathcal{CR}$ gate acting on the next qubit. All rotations corresponding to strings  of the same length $m$ are applied in parallel. 
The transformation at step $m$ is written as
\begin{align}\label{eq:split}
    &\sum_{w\in\{0,1\}^m} c_w\,|w\rangle\,|0\rangle^{\otimes (k-m)}\,|0\rangle\notag\\
    \xrightarrow{\,U_m\,}&
    \sum_{w} c_w\,|w\rangle\,|0\rangle^{\otimes (k-m)}\,|\theta_{w,m}\rangle \notag\\
    \xrightarrow{\,\mathcal{CR}\,}&\;
    \sum_{w} c_w\,|w\rangle\bigl(\cos\theta_{w,m}\,|0\rangle+\sin\theta_{w,m}\,|1\rangle\bigr)\,|0\rangle^{\otimes (k-m-1)}\,|\theta_{w,m}\rangle \notag\\
    \xrightarrow{\,U_m^{\dagger}\,}&\;
    \sum_{w}\bigl(c_{w_0}|w0\rangle+c_{w_1}|w1\rangle\bigr)\,|0\rangle^{\otimes (k-m-1)}\,|0\rangle.
\end{align}
Here, $U_m$ denote an oracle that computes $\theta_{w,m}$ into an angle register,
which is uncomputed after applying \(\mathcal{CR}\) gate.  Iterating the transformation in Equation~\eqref{eq:split} for $m=0,1,\dots,k-1$ yields the desired state 
\begin{equation}
    |\psi_0\rangle=\frac{1}{C}\sum_{n=n_0}^{n_1-1} n^{-\beta/2}|n-n_0\rangle.
\end{equation}
Thus, the central task reduces to constructing $U_m$,  defined as follows.\qed
\begin{definition}[Angle–preparation oracle \(U_m\)]\label{def:Um}
The gate \(U_m\) implements the transformation
\begin{equation}
\sum_{w\in \{0,1\}^m}\alpha_w|w\rangle|0\rangle\mapsto\sum_{w\in \{0,1\}^m}\alpha_w|w\rangle|\theta_{w,m}\rangle,
\end{equation}
where the rotation angle \(\theta_{w,m}\) satisfies
\begin{equation}
\tan^2(\theta_{w,m})=\frac{c^2_{w1}}{c^2_{w0}}
=\frac{\sum\limits_{i=0}^{2^{k-m-1}-1}(2^{k-m-1}(2w+1)+i+n_0)^{-\beta}}
{\sum\limits_{i=0}^{2^{k-m-1}-1}(2^{k-m-1}(2w)+i+n_0)^{-\beta}}
=\frac{S(2^{k-m-1}(2w+1)+n_0,\,2^{k-m-1}(2w+2)-1+n_0,\,\beta)}
{S(2^{k-m-1}(2w)+n_0,\,2^{k-m-1}(2w+1)-1+n_0,\,\beta)},
\end{equation}
with
\begin{equation}
S(a,b,\beta)=\sum_{n=a}^b n^{-\beta}.
\end{equation}
Equivalently, 
\begin{align}\label{eq:theta}
    \theta_{w,m}=&\arctan\Bigl(\sqrt{\frac{S(2^{k-m-1}(2w+1)+n_0,2^{k-m-1}(2w+2)-1+n_0,\beta)}{S(2^{k-m-1} (2w)+n_0,2^{k-m-1}(2w+1)-1+n_0,\beta)}}\Bigr)\notag\\
    =  &\arcsin\Bigl(\sqrt{\frac{S(2^{k-m-1}(2w+1)+n_0,2^{k-m-1}(2w+2)-1+n_0,\beta)}{S(2^{k-m-1}(2w)+n_0,2^{k-m-1}(2w+2)-1+n_0,\beta)}}\Bigr).
\end{align}
\end{definition}
\noindent\textit{Notation.} In practice, \(U_m\) computes an approximation \(\widetilde{\theta}_{w,m}\), which results in an error in the state preparation. The deviation between the ideal and approximate states at step $m$ is bounded as follows:
\begin{align}
&\left\| \sum_w c_w |w\rangle \left( \cos (\theta_{w,m}) |0\rangle + \sin (\theta_{w,m}) |1\rangle \right) |0\rangle^{\otimes (k-m-1)} |0\rangle - \sum_w c_w |w\rangle \left( \cos (\widetilde{\theta}_{w,m}) |0\rangle + \sin (\widetilde{\theta}_{w,m}) |1\rangle \right) |0\rangle^{\otimes (k-m-1)} |0\rangle \right\| \notag \\
=& \left\| \sum_w c_w |w\rangle \left[ ( \cos (\widetilde{\theta}_{w,m} )- \cos (\theta_{w,m}) ) |0\rangle + ( \sin (\widetilde{\theta}_{w,m}) - \sin (\theta_{w,m}) ) |1\rangle \right] \right\| \notag \\
=& \sqrt{\sum_w c_w^2 \cdot 4 \sin^2 \left( \frac{\widetilde{\theta}_{w,m} - \theta_{w,m}}{2} \right)} \notag \\
\leq & \max_w |\widetilde{\theta}_{w,m} - \theta_{w,m}|.
\end{align}
To achieve overall precision \(\varepsilon\), each \(\theta_{w,m}\) must be computed with an error of at most \(\varepsilon/k\) over all $m$. The \(\mathcal{CR}\) gate is decomposed into controlled-\(R_x\) rotations conditioned on the qubits encoding \(\theta_{w,m}\), followed by a phase gate, requiring \(\mathcal{O}(\log(k/\varepsilon))\) gates and no ancilla qubits. Therefore, the dominant resource cost comes from implementing $U_m$ to an accuracy of \(\varepsilon/k\).
\begin{theorem}[Implementation of oracle $U_m$ (Methods, Theorem~2)]\label{the:Um}
    The gate $U_m$, as defined in Definition \ref{def:Um}, can be implemented    such that each rotation angle is estimated to within error $\varepsilon/k$, using a number of gates and ancilla qubits bounded by
    \begin{equation}
    \mathrm{Poly}\Bigl(\log(1/\varepsilon),\,\log(n_1),\,\log\tfrac{1}{|1-\beta|},\,\beta,\,v\Bigr),
    \end{equation}
   where \( v \) denotes the number of significant digits used to represent \( \beta \).
\end{theorem}
\noindent\textit{Proof.} The angle \( \theta_{w,m} \), represented with \( \log(k/\varepsilon) \) qubits, is determined using a bisection procedure applied iteratively on each qubit. For a trial angle $\theta_g$, we compare
    \begin{align}
        &\sin^2(\theta_g)S(2^{k-m-1}2w+n_0,2^{k-m-1}(2w+2)-1+n_0,\beta), \label{eq:p1}\\
     \mathrm{with} \quad   &S(2^{k-m-1}(2w+1)+n_0,2^{k-m-1}(2w+2)-1+n_0,\beta). \label{eq:p2}
    \end{align} 
    This comparison decides whether $\theta_{w,m}>\theta_g$. After $\log(k/\varepsilon)$ comparison,  $\theta_{w,m}$ is computed to within error $\varepsilon/k$. Define
    \begin{equation}
    d=\sin^2(\theta_g)S(2^{k-m-1}\cdot 2w+n_0,\;2^{k-m-1}(2w+2)-1+n_0,\beta)
    -S(2^{k-m-1}(2w+1)+n_0,\;2^{k-m-1}(2w+2)-1+n_0,\beta).
    \end{equation}
    Since $S(a,b,\beta)\ge n_1^{-\beta}$ and relevant angles $\theta$ satisfy 
    $\theta+\frac{\varepsilon}{k}<\frac{\pi}{4}$, consider two angles differing by at least \( \varepsilon/k \). The difference in \( \sin^2 \theta \):
    \begin{equation}
   \sin^2(\theta+\tfrac{\varepsilon}{k})-\sin^2(\theta)=(\sin(\theta+\tfrac{\varepsilon}{k})-\sin(\theta))(\sin(\theta+\tfrac{\varepsilon}{k})+\sin(\theta))>\tfrac{\sqrt{2}\varepsilon}{2k}\sin(\tfrac{\varepsilon}{k})>\tfrac{\varepsilon^2}{2k^2}.
    \end{equation}
    Thus, angles differing by \( \varepsilon/k \) yield a difference in \( d \) of at least $\frac{\varepsilon^2}{2k^2}n_1^{-\beta}$.
    To distinguish such angles correctly, the  error in $d$ must be less than $\frac{\varepsilon^2}{2k^2}n_1^{-\beta}$. Since \( S(a, b, \beta) \leq n_1 \), requirements:
    \begin{center}
        \textbf{(1)}    computing \( S \) to precision \( \frac{\varepsilon^2 n_1^{-\beta}}{12k^2} \),\qquad \qquad\textbf{(2)} \( \sin^2 \theta_g \) to precision \( \frac{\varepsilon^2 n_1^{-1-\beta}}{12k^2} \),
    \end{center}
    ensures the total error meets this requirement.\par
    \textbf{Step~1: Calculation of \boldmath$\sin^2(\theta)$.}  
   Using the expansion
\begin{equation}\label{eq:sin}
\sin^2(\theta)=\frac{1-\cos(2\theta)}{2}
=\sum_{n=1}^\infty d_n\theta^{2n},\qquad d_n=\frac{(-1)^{n+1}2^{2n}}{2(2n)!},
\end{equation}
we approximate \( \sin^2 (\theta) \) for \( \theta < \frac{\pi}{4} \). Truncating after \( l_2 > 5 \) terms, the error is:
\begin{equation}
    \frac{(\frac{\pi}{2})^{2l_2+2}}{2(2l_2+2)!}<\frac{(\frac{\pi}{2})^{2l_2+2}}{(\frac{2l_2+2}{e})^{2l_2+2}}=(\frac{e\pi}{4l_2+4})^{2l_2+2}<(\frac{1}{2})^{l_2},
\end{equation} 
using Stirling’s approximation. To achieve precision \( \frac{\varepsilon^2 n_1^{-1-\beta}}{24k^2} \), we choose:
\begin{equation}
l_2 = 2\log\Bigl(\frac{k}{\varepsilon}\Bigr) + (1+\beta)\log(n_1)+5.
\end{equation}
Storing coefficients \( d_n \) to precision \( \frac{\varepsilon^2 n_1^{-1-\beta}}{96k^2} \) ensures that the rounding error is bounded by:
\begin{equation}
    \frac{\varepsilon^2n_1^{-1-\beta}}{96k^2}(\frac{\pi}{4}+(\frac{\pi}{4})^2+\cdots)<\frac{\varepsilon^2n_1^{-1-\beta}}{24k^2}.
\end{equation}
Thus, the total truncation and rounding error is within $\frac{\varepsilon^2 n_1^{-1-\beta}}{12k^2}$, satisfying the requirement \textbf{(1)}.\par
The expression approximating \( \sin^2(\theta) \) is a polynomial of degree \( 2l_2 \). The parameter \( \theta \) is encoded using \( \log\left(\frac{k}{\varepsilon}\right) \) qubits, the coefficients require \( \log\left(\frac{\varepsilon^2 n_1^{-1-\beta}}{96k^2}\right) \) qubits, and the output requires \( \frac{\varepsilon^2 n_1^{-1-\beta}}{12k^2} \) qubits.
By Lemma~\ref{polynomial}, the gate complexity is
 \begin{equation}\label{step1,1}
     4\log^2(\frac{k}{\varepsilon})l_2^2+4\log(\frac{k}{\varepsilon})\log(\frac{96k^2n_1^{1+\beta}}{\varepsilon^2})l_2^2+2l_2\log(\frac{12k^2n_1^{1+\beta}}{\varepsilon^2})=\mathrm{Poly}(\log(n_1),\log(1/\varepsilon),\beta),
 \end{equation}
and the ancilla qubit count is
\begin{equation}\label{step1,2}
4l_2\log(\frac{k}{\varepsilon})+\log(\frac{96k^2n_1^{1+\beta}}{\varepsilon^2})=\mathrm{Poly}(\log(n_1),\log(1/\varepsilon),\beta).
\end{equation}
\par
\textbf{Step~2: Estimation of \boldmath$S$.} 
We will prove Lemma~\ref{S oracle}, which states that the partial sum \( S(a, b, \beta) \) can be computed to precision \( \epsilon \) with:
\begin{equation}\label{step2}
\mathrm{Poly}\left( \log(1/\epsilon), \log n_1, \log \left( \frac{1}{|1 - \beta|} \right),v \right)
\end{equation}
gates and ancilla qubits,  where \( v \) denotes the significant digits of \( \beta \). To satisfy requirement \textbf{(2)}, we set \( \epsilon = \frac{\varepsilon^2 n_1^{-\beta}}{12k^2} \) and the complexity becomes:
\[
\mathrm{Poly}\left( \log \left( \frac{12k^2 n_1^{\beta}}{\varepsilon^2} \right), \log (n_1), \log \left( \frac{1}{|1 - \beta|} \right),v \right) = \mathrm{Poly}\left( \log(1/\varepsilon), \log (n_1), \log \left( \frac{1}{|1 - \beta|} \right), \beta,v\right).
\] \par
Conclusively, we give a total count of required resourced for constructing $U_m$. The bisection procedure requires \( \lceil \log(k/\varepsilon) \rceil \) comparisons to compute \( \theta_{w,m} \) to precision \( \varepsilon/k \). Each comparison involves two evaluations of \( S \) (for the partial sums in \eqref{eq:p1} and \eqref{eq:p2}) and one evaluation of \( \sin^2 \theta_g \). Thus, the total resource cost for \( U_m \) is multiplied by a factor of \( \mathcal{O}(\log(k/\varepsilon)) \), which is absorbed into the overall polynomial complexity. Combining the resource costs from Steps 1 and 2 (concluded in Equation \eqref{step1,1},\eqref{step1,2},\eqref{step2}), the total gate and ancilla qubit complexity for implementing the oracle \( U_m \) is bounded by
\begin{equation}
    \mathrm{Poly}\Bigl(\log(1/\varepsilon),\,\log(n_1),\,\log\tfrac{1}{|1-\beta|},\,\beta,\,v\Bigr),
    \end{equation}
 as stated. As a preliminary step of Lemma~\ref{S oracle}, an approximation of \( S \) is introduced in Proposition~\ref{lm:split}. \qed
\begin{proposition}[Euler–Maclaurin approximation of $S$]
    \label{lm:split}
For $\beta>0$, $\beta\neq1$, and integers $a\leq b$, there exists an approximation $\widetilde{S(a, b, \beta)}$ to the partial sum
$S(a, b, \beta) = \textstyle\sum_{n=a}^{b} n^{-\beta}$, such that
\begin{equation}
\left| S(a, b, \beta) - \widetilde{S(a, b, \beta)} \right| < \frac{\epsilon}{2},
\end{equation}
where
\begin{equation}\label{eq:widetildeS}
    \widetilde{S(a,b,\beta)}=\frac{1}{1-\beta}(b^{1-\beta}-a^{1-\beta})+ \frac{a^{-\beta}+b^{-\beta}}{2}-\sum_{r=1}^{l_3}\frac{B_{2r}\Gamma(\beta+2r-1)}{(2r)!\,\Gamma(\beta)}( b^{-\beta-2r+1}-a^{-\beta-2r+1}),
\end{equation}
with $l_3=\lceil\frac{1}{2}log_{2\pi}\frac{8}{\epsilon}\rceil$, provided $a>\lceil \beta+2m\rceil$.
\end{proposition}
\noindent\textit{Proof.} Recall the Euler–Maclaurin summation formula:
\begin{equation}
\sum_{k=a}^{b} f(k)
  = \int_{a}^{b} f(x)\,\mathrm{d}x
  + \frac{f(a)+f(b)}{2}
  + \sum_{r=1}^{l_3} \frac{B_{2r}}{(2r)!}\Bigl(f^{(2r-1)}(b)-f^{(2r-1)}(a)\Bigr)
  + R_{l_3},
\end{equation}
where $B_{2r}$ are Bernoulli Numbers, and the remainder is given by
\begin{equation}
R_{l_3}= \frac{(-1)^{l_3+1}}{(2l_3)!}\int_{a}^{b} B_{2l_3}(x-\lfloor x\rfloor)\,
        f^{(2l_3)}(x)\,\mathrm{d}x.
\end{equation}
Setting $f(x)=x^{-\beta}$, we have
\begin{align}
    S(a,b,\beta)
    =&\int_{a}^{b} x^{-\beta}\,\mathrm{d}x
  + \frac{a^{-\beta}+b^{-\beta}}{2}
  + \sum_{r=1}^{l_3} \frac{B_{2r}}{(2r)!}(-1)^{2r-1}\Bigl(\prod\limits_{i=0}^{2r-2}(\beta+i)\Bigr)( b^{-\beta-2r+1}-a^{-\beta-2r+1})+ R_{l_3}\notag\\
  =&\frac{1}{1-\beta}(b^{1-\beta}-a^{1-\beta})+ \frac{a^{-\beta}+b^{-\beta}}{2}-\sum_{r=1}^{l_3}\frac{B_{2r}\Gamma(\beta+2r-1)}{(2r)!\,\Gamma(\beta)}( b^{-\beta-2r+1}-a^{-\beta-2r+1})+ R_{l_3},
\end{align}
in which the reminder term is given by
\begin{equation}
    R_{l_3}=\frac{(-1)^{l_3+1}}{(2l_3)!}\int_{a}^{b} B_{2l_3}(x-\lfloor x\rfloor)\,
        \frac{\Gamma(\beta+2l_3)}{\Gamma(\beta)}x^{-\beta-2l_3}\,\mathrm{d}x.
\end{equation}
Using the bound on Bernoulli numbers:
\begin{equation}
    \frac{2(2l_3)!}{(2\pi)^{2l_3}}\frac{1}{1-2^{-2l_3}}< |B_{2l_3}|=\frac{2(2l_3)!}{(2\pi)^{2l_3}}\zeta(2l_3)<\frac{2(2l_3)!}{(2\pi)^{2l_3}}\frac{1}{1-2^{1-2l_3}},
\end{equation}
we obtain
\begin{align}
    |R_{l_3}|&\leq\frac{|B_{2l_3}|}{(2l_3)!}\beta(\beta+1)\cdots (\beta+2l_3-1)\int_{a}^{b}x^{-\beta-2l_3}\,dx \notag\\
    &=\left|\frac{\beta(\beta+1)\cdots(\beta+2l_3-1)B_{2l_3}(a^{-\beta-2l_3+1}-b^{-\beta-2l_3+1})}{(2l_3)!(\beta+2l_3-1)}\right|\notag\\
    &<\frac{\beta(\beta+1)\cdots(\beta+2l_3-2)2a^{-\beta-2l_3+1}}{(2\pi)^{2l_3}(1-2^{1-2l_3})}\notag\\
    &<\frac{4}{(2\pi)^{2l_3}}\frac{\beta(\beta+1)\cdots(\beta+2l_3-2)}{a^{\beta+2l_3-1}}.  
\end{align}
Thus, choosing $l_3=\lceil\frac{1}{2}\log_{2\pi}\frac{8}{\epsilon}\rceil$ guarantees $|R_{l_3}|<\frac{\epsilon}{2}$, provided $a>\lceil\beta+2l_3\rceil$.  \qed
\begin{corollary}
The Bernoulli numbers satisfy $B_{0}=1$, and for all $q\geq 1$,
\begin{equation}
\sum_{p=0}^{q}\binom{q+1}{p}B_{p}=0.
\end{equation}
\end{corollary}
\noindent The Bernoulli numbers can be computed efficiently using the Akiyama–Tanigawa algorithm~\cite{akiyama2001multiple}, with a complexity $O(p^{2})$ for computing $B_p$.
\begin{lemma}[$\widehat{S(a,b,\beta)}$ calculation oracle]\label{S oracle}
Let $\mathcal{S}(\beta)$ be an oracle implementing 
\begin{equation}
\sum_{n_0\leq a\leq b\leq n_1-1}\alpha_{ab}|a\rangle|b\rangle|0\rangle\mapsto\sum_{n_0\leq a\leq b\leq n_1-1}\alpha_{ab}|a\rangle|b\rangle|\widehat{S(a,b,\beta)}\rangle,
\end{equation}
where $\widehat{S(a,b,\beta)}$ is an $\frac{\epsilon}{2}$-approximation of $\widetilde{S(a,b,\beta)}$  as defined in Equation \eqref{eq:widetildeS}, and therefore an $\epsilon$-approximation of $S(a,b,\beta)$.
Then $\mathcal{S}(\beta)$ can be constructed  using
    \begin{equation}
        \mathrm{Poly}(\log(1/\epsilon),\log(n_1),\log(\tfrac{1}{|1-\beta|}),v)
    \end{equation}
gates and ancilla qubits, where $l_3=\lceil\frac{1}{2}\log_{2\pi}\frac{8}{\epsilon}\rceil$, $n_0>\beta+2l_3$, and $v$ is the number of significant digits of $\beta$.
\end{lemma}
\noindent\textit{Proof sketch.}
We compute \(\widetilde{S(a, b, \beta)}\), as defined in Equation \eqref{eq:widetildeS}, with an error of at most \(\epsilon/2\) by evaluating its exponential terms and finite Bernoulli-series sums:
\begin{align}
    \widetilde{S(a,b,\beta)}=&b^{-\beta-2l_3+1}(\frac{1}{1-\beta}b^{2l_3}+\frac{1}{2}b^{2l_3-1}-\sum_{r=1}^{l_3} \frac{B_{2r}\Gamma(\beta+2r-1)}{(2r)!\,\Gamma(\beta)} b^{2l_3-2r})\notag\\
    -&a^{-\beta-2l_3+1}(\frac{1}{1-\beta}a^{2l_3}-\frac{1}{2}a^{2l_3-1}-\sum_{r=1}^{l_3} \frac{B_{2r}\Gamma(\beta+2r-1)}{(2r)!\,\Gamma(\beta)} a^{2l_3-2r})\notag\\
    =& b^{-\beta-2l_3+1}\,P_b(\beta)- a^{-\beta-2l_3+1}\,P_a(\beta),
\end{align}
where
\[
P_b(\beta)=\frac{1}{1-\beta}b^{2l_3}+\frac{1}{2}b^{2l_3-1}
  -\sum_{r=1}^{l_3} \frac{B_{2r}\Gamma(\beta+2r-1)}{(2r)!\,\Gamma(\beta)} b^{2l_3-2r},
\]
\[
P_a(\beta)=\frac{1}{1-\beta}a^{2l_3}-\frac{1}{2}a^{2l_3-1}
  -\sum_{r=1}^{l_3} \frac{B_{2r}\Gamma(\beta+2r-1)}{(2r)!\,\Gamma(\beta)} a^{2l_3-2r}.
\]
Define the \( r \)-th term in the sum of \( P_b(\beta) \):
\[
T_b(r)=\frac{B_{2r}\,\Gamma(\beta+2r-1)}{(2r)!\,\Gamma(\beta)}\,b^{2l_3-2r}.
\]
Then the ratio of consecutive terms is
\[
\left|\frac{T_b(r+1)}{T_b(r)}\right|
= \left|\frac{B_{2r+2}}{B_{2r}}\right|
  \frac{(\beta+2r-1)(\beta+2r)}{(2r+2)(2r+1)}\frac{1}{b^{2}}.
\]
Using the bounds of Bernoulli number
\begin{equation}
    \frac{2(2r)!}{(2\pi)^{2r}}\frac{1}{1-2^{-2r}}< |B_{2r}|=\frac{2(2r)!}{(2\pi)^{2r}}\zeta(2r)<\frac{2(2r)!}{(2\pi)^{2r}}\frac{1}{1-2^{1-2r}},
\end{equation}
for \( b > \beta + 2 l_3 \), we obtain:
\[
\left|\frac{T_b(r+1)}{T_b(r)}\right| <\frac{(\beta+2r-1)(\beta+2r)\bigl(1-2^{-2r}\bigr)}{(2\pi)^2b^2\bigl(1-2^{1-2(r+1)}\bigr)} \
< \frac{1}{2\pi^2}.
\]
Thus the Bernoulli series decreases geometrically, 
\begin{equation}
\left|\sum_{r=1}^{l_3} \frac{B_{2r}\Gamma(\beta+2r-1)}{(2r)!\Gamma(\beta)}b^{2l_3-2r}\right|
< \left|\frac{B_2\beta b^{2l_3-2}}{2\left(1-\frac{1}{2\pi^2}\right)}\right|
< \frac{\beta b^{2l_3-2}}{6}
< \frac{b^{2l_3-1}}{2}.
\end{equation}
By combining the fact that $\frac{b^{2l_3-1}}{2}<\frac{b^{2l_3}}{2|1-\beta|}$, we obtain
\begin{equation}
    |P_b(\beta)|<\frac{2b^{2l_3}}{|1-\beta|},
\end{equation}
and similarly,
\begin{equation}
    |P_a(\beta)|<\frac{2a^{2l_3}}{|1-\beta|}.
\end{equation}
Hence, to achieve overall precision~$\frac{\epsilon}{2}$, it suffices to:
\begin{enumerate}
  \item Compute $a^{-\beta-2l_3+1}$ and $b^{-\beta-2l_3+1}$ to precisions:
  \begin{equation}
      \frac{\epsilon |1-\beta|}{16{a^{2l_3}}}\qquad\text{and}\qquad\frac{\epsilon |1-\beta|}{16{b^{2l_3}}},
  \end{equation}
 respectively. We will prove in Proposition~\ref{prop:endpoint} that these require at most 
 \begin{equation}
     \mathrm{Poly}(\log(n_1),\log(\tfrac{1}{|1-\beta|}),\log(1/\epsilon),v)
 \end{equation}
 gates and ancilla qubits.
\item Compute the  polynomial–Bernoulli sums $P_a(\beta)$ and $P_b(\beta)$ to precision $\frac{\epsilon}{8}$, since $a^{-\beta-2l_3+1}$ and $b^{-\beta-2l_3+1}$ are both smaller than 1.  We will prove in Proposition ~\ref{prop:summation} that these require:
\begin{equation}
    \mathrm{Poly}(\log(1/\epsilon),\log(n_1),\log(\tfrac{1}{|1-\beta|}),v)
\end{equation}
gates and ancilla qubits.  
\item Perform final multiplications and additions to combine terms. As the exponential terms and Bernoulli-series sums require at most
\[
    \log(\frac{16n_1^{2l_3}}{|1-\beta|\epsilon}),\qquad \log(\frac{2n_1^{2l_3}}{|1-\beta|})+\log(\frac{8}{\epsilon})
\]
significant qubits, respectively. 
\end{enumerate}
Conclusively, the total number of required gates and ancilla qubits can be bounded by 
\begin{equation}
\mathrm{Poly}(\log(1/\epsilon),\log(n_1),\log(\tfrac{1}{|1-\beta|}),v),
\end{equation}
given that $l_3=\lceil\frac{1}{2}\log_{2\pi}\frac{8}{\epsilon}\rceil$. Hence the Lemma follows. Next, we  detail Proposition~\ref{prop:endpoint} and Proposition~\ref{prop:summation}.  \qed
\begin{proposition}~\label{prop:endpoint}
    The oracles
    \begin{equation}
        \sum_{b=n_0}^{n_1-1}\alpha_b|b\rangle|0\rangle\mapsto \sum_{b=n_0}^{n_1-1}\alpha_b|b\rangle |\widetilde{b^{-\beta-2l_3+1}}\rangle \qquad\mathrm{and}\qquad \sum_{a=n_0}^{n_1-1}\alpha_a|a\rangle|0\rangle\mapsto \sum_{a=n_0}^{n_1-1}\alpha_a|a\rangle |\widetilde{a^{-\beta-2l_3+1}}\rangle
    \end{equation}
    can be constructed with error bounds
    \begin{equation}
      |\widetilde{b^{-\beta-2l_3+1}}-b^{-\beta-2l_3+1}|<\frac{\epsilon |1-\beta|}{16{b^{2l_3}}}, \qquad|\widetilde{a^{-\beta-2l_3+1}}-a^{-\beta-2l_3+1}|<\frac{\epsilon |1-\beta|}{16{a^{2l_3}}},
  \end{equation}
 using
    \begin{equation}
            \mathrm{Poly}(\log(n_1),\log(\tfrac{1}{|1-\beta|}),\log(1/\epsilon),v)
    \end{equation}
    gates and ancilla qubits, where $n_0>\beta+2l_3$, and $v$ denotes the number of significant digits of $\beta$.
\end{proposition}
\noindent\textit{Proof.} We focus on constructing the oracle for \( \widetilde{b^{-\beta - 2 l_3 + 1}} \); the case for \( a \) is analogous.  W riting \( b^{-\beta - 2 l_3 + 1} = 2^{-(\beta + 2 l_3 - 1) \log b} \) , the computation proceeds by approximating \( \log b \), multiplying by \( -(\beta + 2 l_3 - 1) \), and exponentiating. We require the approximation errors from the logarithm and from the exponentiation to be both bounded by $\frac{\epsilon |1-\beta|}{32{b^{2l_3}}}$, so that the total error is bounded by $\frac{\epsilon |1-\beta|}{16{b^{2l_3}}}$.\par
\textbf{Step~1: Approximating \boldmath$\log(b)$.} 
If $(\beta+2
l_3-1)|\widetilde{\log(b)}-\log(b)|<\frac{1}{2}$, then
\begin{align}
&\bigl|2^{-(\beta+2l_3-1)\log(b)}-2^{-(\beta+2l_3-1)\widetilde{\log(b)}}\bigr| \notag\\
&=2^{-(\beta+2l_3-1)\log(b)}
\Bigl|1-2^{-(\beta+2l_3-1)(\widetilde{\log(b)}-\log(b))}\Bigr| \notag\\
&< b^{-\beta-2l_3+1}(\beta+2l_3-1)\,|\widetilde{\log(b)}-\log(b)|,
\end{align}
since \( |1 - 2^x| \leq |x| \) for \( |x| \leq\frac{1}{2}\). To bound this error by $\frac{\epsilon|1-\beta|}{32b^{2l_3}}$,
it suffices to compute $\log(b)$ to precision
\begin{equation}
\frac{\epsilon|1-\beta|}{32b^{2l_3}}\frac{b^{\beta+2l_3-1}}{(\beta+2l_3-1)}
=\frac{b^{\beta-1} \epsilon |1-\beta|}{32(\beta+2l_3-1)}>\frac{1}{32}n_1^{-2}|1-\beta|\,\epsilon,
\end{equation}
which satisfies $\frac{1}{32}n_1^{-2}|1-\beta|\,\epsilon(\beta+2l_3-1)<\frac{1}{32}\epsilon<\frac{1}{2}$. Therefore, we set the required qubit of precision as
\begin{equation}
    p_1=\log(32n_1^{2}\tfrac{1}{|1-\beta|}\frac{1}{\epsilon})=5+2\log(n_1)+\log(\tfrac{1}{|1-\beta|})+\log(\tfrac{1}{\epsilon}).
\end{equation}
By Lemma \ref{the:log}, computing $\log(b)$  to $p_1$ digits of precision requires
\begin{center}
    $\mathcal{O}(\log(n_1)^3p_1^2)$ gates and $\mathcal{O}(\log(n_1)^2p_1)$ ancilla qubits.
\end{center}\par
\textbf{Step~2: Multiplication with \boldmath$\beta+2l_3-1$.} 
Multiplying \( \log b \) by \( \beta + 2 l_3 - 1 \), which are represented to at most \( \log (n_1) + p_1 \) and \( \log (n_0) + v \) bits respectively, requires:
 \begin{center}
 $(\log(n_1)+p_1)(\log(n_0)+v)$ gates and $(\log(n_1)+p_1)+(\log(n_0)+v)$ ancilla qubits,
 \end{center}
 where $v$ is the number of significant digits of $\beta$.\par
 \textbf{Step~3: Exponentiation.} 
Decompose
\begin{equation}
    b^{-\beta-2l_3+1}=2^{\lfloor -(\beta+2l_3-1)\log(b)\rceil}\cdot2^{\{-(\beta+2l_3-1)\log(b)\}},
\end{equation}
where $\lfloor y \rceil$ denote the nearest integer of $y$ and $\{y\}= y - \lfloor y \rceil$  denote the fractional part. The fractional exponential is evaluated by:
\begin{equation}
    2^x=\mathrm{e}^{\ln2\cdot x}=\sum_{n=0}^{\infty}\frac{(\ln2)^n}{n!}x^n,\qquad-\frac{1}{2}\leq x<\frac{1}{2}.
\end{equation}
To bound the total error by $\frac{\epsilon|1-\beta|}{32b^{2l_3}}$, 
we target an error of
\begin{equation}
\frac{\epsilon|1-\beta|}{32b^{2l_3}}\frac{1}{\sqrt{2}b^{-(\beta+2l_3-1)}}=\frac{b^{\beta+2l_3-1}\epsilon|1-\beta|}{32\sqrt{2}b^{2l_3}}
>\frac{1}{64}b^{\beta-1}|1-\beta|\epsilon
>\frac{1}{64}n_1^{-1}|1-\beta|\epsilon.
\end{equation}
in fractional exponential. Let 
\begin{equation}
  l_4=\log(64n_1\tfrac{1}{|1-\beta|}\frac{1}{\epsilon})=6+\log(n_1)+\log(\tfrac{1}{|1-\beta|})+\log(\tfrac{1}{\epsilon}).
\end{equation}
Because $|x|\leq\frac{1}{2}$, truncating the exponential series at degree $l_4$ gives an error
\begin{equation}
    \frac{2(\ln 2/2)^{l_4}}{(l_4)!}
    <\frac{1}{2^{l_4+1}}=\frac{1}{128}n_1^{-1}|1-\beta|\epsilon.
\end{equation}
With coefficients stored with $l_4+1$ fractional qubits,  the rounding error is at most 
\begin{equation}
    \frac{1}{128}n_1^{-1} |1-\beta|\epsilon(1+\frac{1}{2}+\frac{1}{4}+\cdots)\leq\frac{1}{128}n_1^{-1} |1-\beta|\epsilon\leq \frac{1}{128}n_1^{\beta-1} |1-\beta|\epsilon,
\end{equation}
yielding a total error  at most $\frac{1}{64}n_1^{-1}|1-\beta|\epsilon$.\par
 By Lemma~\ref{polynomial}, evaluating the degree-$l_4$ polynomial in the argument
 $\{-(\beta+2l_3-1)\log(b)\}$, which is represented with at most $(p_1+\log(n_1)+\log(n_0)+v)$ bits through  multiplication, and with coefficients  stored with $l_4+1$ qubits, to achieve precision $\frac{\epsilon|1-\beta|}{32b^{2l_3}}$ requires
 \begin{align}
     &l_4^2(p_1+\log(n_0)+\log(n_1)+v)^2+l_4^2(p_1+\log(n_0)+\log(n_1)+v)(l_4+1)
     +\log(\frac{32b^{2l_3}}{|1-\beta|\epsilon})l_4\notag\\
     =&\mathrm{Poly}(\log(n_1),\log(\tfrac{1}{|1-\beta|}),\log(\tfrac{1}{\epsilon}),v)
 \end{align}
 gates, and 
  \begin{equation}
     2(p_1+\log(n_0)+\log(n_1)+v)^2l_4+l_4+1
     =\mathrm{Poly}(\log(n_1),\log(\tfrac{1}{|1-\beta|}),\log(\tfrac{1}{\epsilon}),v)
 \end{equation}
 ancilla qubits. Thus, combine the three parts and the claimed resource bounds follow. \qed
 \begin{proposition}\label{prop:summation}
     The oracles
    \begin{equation}
       \sum_{b=n_0}^{n_1-1}\alpha_b |b\rangle|0\rangle\mapsto \sum_{b=n_0}^{n_1-1}\alpha_b|b\rangle |P_b(\beta)\rangle \qquad\mathrm{and} \qquad \sum_{a=n_0}^{n_1-1}\alpha_a|a\rangle|0\rangle\mapsto \sum_{a=n_0}^{n_1-1}\alpha_a|a\rangle |P_a(\beta)\rangle,
    \end{equation}
where
\begin{align}\label{type2}
P_b(\beta)&=\frac{1}{1-\beta}b^{2l_3}+\frac{1}{2}b^{2l_3-1}
  -\sum_{r=1}^{l_3} \frac{B_{2r}\Gamma(\beta+2r-1)}{(2r)!\,\Gamma(\beta)} b^{2l_3-2r},\notag\\
P_a(\beta)&=\frac{1}{1-\beta}a^{2l_3}-\frac{1}{2}a^{2l_3-1}
  -\sum_{r=1}^{l_3} \frac{B_{2r}\Gamma(\beta+2r-1)}{(2r)!\,\Gamma(\beta)} a^{2l_3-2r},
\end{align}
can be constructed to precision $\frac{\epsilon}{8}$ using
\begin{equation}
    \mathrm{Poly}(\log(1/\epsilon),\log(n_1),\log(\tfrac{1}{|1-\beta|}))
\end{equation}
gates and ancilla qubits, in which $n_0>\beta+2l_3$.
 \end{proposition}
\noindent\textit{Proof.} We focus on constructing the oracle for $P_b(\beta)$, as the case for $P_a(\beta)$ is analogous. All the coefficients are stored with precision $\frac{\epsilon}{16n_1^{2l_3}}$, so the total rounding  is bounded by
\begin{equation}
    \frac{\epsilon}{16n_1^{2l_3}}(b^{2l_3}+b^{2l_3-1}+\cdots)<\frac{\epsilon}{16n_1^{2l_3}}2b^{2l_3}\leq\frac{\epsilon}{8}.
\end{equation}
Hence the required number of qubit for each coefficient is at most 
\begin{equation}
    p_2=\log(\frac{16n_1^{2l_3}}{\epsilon})=\log(1/\epsilon)+4+2l_3\log(n_1).
\end{equation}
The polynomial in Equation \eqref{type2} has degree $2l_3$, with input $b$ requiring at most $\log(n_1)$ qubits, and the output targeted to precision $\frac{\epsilon}{8}$ with absolute magnitude bounded by $\log(\frac{2}{|1-\beta|}n_1^{2l_3})$. Lemma~\ref{polynomial} implies that this computation can be performed using at most
\begin{align}
    &l_3^2\log^2(n_1)+l_3^2\log(n_1)p_2+l_3\bigl(\log(\frac{2}{|1-\beta|}n_1^{2l_3})+\log(\frac{8}{\epsilon})\bigr)\notag\\
    =&\mathrm{Poly}\Bigl(\log(n_1),\log(\tfrac{1}{|1-\beta|}),\log(\tfrac{1}{\epsilon})\Bigr)
\end{align}
gates, and 
\begin{equation}
    2l_3\log(n_1)+p_2=\mathrm{Poly}(\log(n_1),\log(1/\epsilon))
\end{equation}
ancilla qubits. \qed

\begin{corollary}[Extended state preparation]
    \label{co:extended}

    Let $\beta>0,\;\beta\neq1$. Define 
    \begin{equation}
        |\psi_0\rangle_{\mathrm{e}}=\frac{1}{C}\sum_{n=1}^{n_1-1}n^{-\beta/2}|n\rangle, \qquad C=\sqrt{\sum_{n=1}^{n_1-1}n^{-\beta}},
    \end{equation}
    where $c = \lceil \frac{1}{2} \log_{2\pi} \frac{24}{\varepsilon} \rceil$, $n_0 - 1$ is the smallest power of 2 greater than $\lceil \beta + 2c \rceil$, and $n_1 - n_0$ is a power of 2. Then the state $|\psi_0\rangle_e$ can be prepared on a gate-based quantum computer to precision $\varepsilon > 0$ using
    \begin{equation}
        \mathrm{Poly}(\log(1/\varepsilon),\log(n_1),\log(\tfrac{1}{|1-\beta|}),\beta,v),
    \end{equation}
    gates and ancilla qubits, where $v$ is the number of significant digits used to represent $\beta$.
\end{corollary}
\noindent\textit{Proof.} 
By Theorem \ref{th:trusi}, the truncated state
\begin{equation}
    |\psi_0\rangle=\frac{1}{C}\sum_{n=n_0}^{n_1-1}n^{-\beta/2}|n\rangle
\end{equation}
  can be prepared to precision $\frac{\varepsilon}{3}$ with
 \begin{equation}
     \mathrm{Poly}(\log(1/\varepsilon),\log(n_1),\log(\tfrac{1}{|1-\beta|}),\beta,v)
 \end{equation}
 gates and ancilla qubits, for $c = \lceil \frac{1}{2} \log_{2\pi} \frac{24}{\varepsilon} \rceil$ and $n_0$  greater than $\lceil \beta + 2c \rceil + 1$. Following the approach of direct amplitude splitting~\cite{long2001efficient}, we can also prepare
 \begin{equation}
     |\psi_1\rangle=\frac{1}{C_1}\sum_{n=1}^{n_0-1}n^{-\beta/2}|n\rangle,
 \end{equation}
to precision $\frac{\varepsilon}{3}$, using $\mathrm{Poly}(n_0,\log(1/\varepsilon))$ gates. An auxiliary qubit coherently combines the two states:
\begin{equation}
    \frac{|0\rangle+\gamma|1\rangle}{\sqrt{\gamma^2+1}}|0\rangle
    \rightarrow \frac{|0\rangle|\psi_0\rangle+\gamma|1\rangle|\psi_1\rangle}{\sqrt{\gamma^2+1}}.
\end{equation}
Applying a Hadamard gate gives:
\begin{equation}
    \xrightarrow{H} \frac{(|0\rangle+|1\rangle)|\psi_0\rangle+\gamma(|0\rangle-|1\rangle)|\psi_1\rangle}{\sqrt{2\gamma^2+2}}
    =\frac{|0\rangle (|\psi_0\rangle+\gamma|\psi_1\rangle)+|1\rangle(|\psi_0\rangle-\gamma|\psi_1\rangle)}{\sqrt{2\gamma^2+2}}.
\end{equation}
Here, $\gamma=\frac{C_1}{C}=\sqrt{\frac{S(1,n_0-1,\beta)}{S(n_0,n_1-1,\beta)}}$, where the denominator estimated via Euler-Maclaurin summation method and the numerator is computed directly. If the ancilla is measured in $|0\rangle$, the desired normalized superposition is obtained. If not, a correction step—using an additional register for comparison with $n_0$ and a controlled-$Z$ operation-restores the target state. The comparator writes a 1 in an auxiliary qubit iff the index is smaller than $n_0$. The process is listed as follows:
\begin{equation}
    (|\psi_0\rangle-\gamma|\psi_1\rangle)|0\rangle\xrightarrow{\text{comparator}}|\psi_0\rangle|0\rangle-\gamma|\psi_1\rangle|1\rangle\xrightarrow{\text{controlled-}Z}|\psi_0\rangle|0\rangle+\gamma|\psi_1\rangle|1\rangle\xrightarrow{\text{undo comparator}} (|\psi_0\rangle
+\gamma|\psi_1\rangle)|0\rangle.
\end{equation}
If $\gamma$ is approximated by $\widetilde{\gamma}$, the error is bounded by
\begin{align}
&\left\|\frac{|\psi_0\rangle+\widetilde{\gamma}|\psi_1\rangle}{\sqrt{1+\widetilde{\gamma}^2}}
-\frac{|\psi_0\rangle+\gamma|\psi_1\rangle}{\sqrt{1+\gamma^2}}\right\|
=\sqrt{2-2\frac{1+\gamma\widetilde{\gamma}}{\sqrt{(1+\gamma^2)(1+\widetilde{\gamma}^2)}}} 
=\sqrt{2-2\sqrt{1-\frac{(\gamma-\widetilde{\gamma})^2}{(1+\gamma^2)(1+\widetilde{\gamma}^2)}}} \notag\\
<& \sqrt{2-2\sqrt{1-(\gamma-\widetilde{\gamma})^2}} 
< \sqrt{2|\gamma-\widetilde{\gamma}|}.
\end{align}
Thus, computing $\gamma$ to precision $\frac{\varepsilon^2}{18}$ ensures an additional error at most $\frac{\varepsilon}{3}$. This can be achieved by estimating
 $C$ with error $n_0^{-\beta}\frac{\varepsilon^2}{36}$ and $C_1$ with error $\frac{\varepsilon^2}{36}\frac{C^2}{C_1}>\frac{\varepsilon^2}{36}\frac{n_0^{-2\beta}}{n_0}=\frac{\varepsilon^2}{36}n_0^{-1-2\beta}$ using Euler-Maclaurin summation with $\mathrm{Poly}(\log(1/\varepsilon),\log(\tfrac{1}{|1-\beta|}),\beta,\log(n_1),v)$ classical computations. Hence, $|\psi_0\rangle_{\mathrm{e}}$ can be prepared with accuracy $\varepsilon$ using
\begin{equation}
    \mathrm{Poly}(\log(1/\varepsilon),\log(n_1),\log(\tfrac{1}{|1-\beta|}),\beta,v)
\end{equation}
gates and ancilla qubits. \qed
\begin{theorem}[Initial state preparation (Methods, Theorem~3)]\label{th:initsi}
Let $\beta>0$, $\beta \neq 1$ and $N\in\mathbb{N}$. Define
    \begin{equation}
        |\psi_0\rangle=\frac{1}{C}\sum_{n=1}^{N}n^{-\beta/2}|n\rangle, \qquad C=\sqrt{\sum_{n=1}^{N}n^{-\beta}}.
    \end{equation}
    Then  $|\psi_0\rangle$ can be prepared on a quantum computer to precision $\varepsilon>0$, with success probability at least $(\frac{1}{2}-\frac{\varepsilon}{3})$. The required number of gates and ancilla qubits is bounded by
    \begin{equation}
        \mathrm{Poly}(\log(1/\varepsilon),\log(N),\log(\tfrac{1}{|1-\beta|}),\beta,v),
    \end{equation}
where $v$ denotes the number of significant digits used to represent $\beta$.
\end{theorem}
\noindent\textit{Proof.} 
Set
$c = \Biggl\lceil \frac{1}{2}\log_{2\pi}\!\Bigl(\frac{144}{\varepsilon}\Bigr)\Biggr\rceil$, and choose $n_0$ so that $(n_0-1)$ is the smallest power of 2 larger than $\lceil \beta+2c\rceil$.  
Let $n_1$  be the smallest integer larger than $N$ with $n_1-n_0$ a power of $2$.  By Corollary \ref{co:extended}, the extended state
\begin{equation}
    |\psi_0\rangle=\frac{1}{C}\sum_{n=1}^{n_1-1}n^{-\beta/2}|n\rangle
\end{equation}
can be constructed to precision $\tfrac{\varepsilon}{6}$ using 
\begin{equation}
    \mathrm{Poly}(\log(1/\varepsilon),\log(n_1),\log(\tfrac{1}{|1-\beta|}),\beta,v)=\mathrm{Poly}(\log(1/\varepsilon),\log(N),\log(\tfrac{1}{|1-\beta|}),\beta,v)
\end{equation}
gates and ancilla qubits. \par
Let $P$ denote the projector onto the subspace spanned by $\{\ket{n}:1\le n\le N\}$, given by:
\begin{equation}
    P=\sum_{n=1}^{N}|n\rangle\langle n|,
\end{equation}
which can be implemented using an ancilla qubit, a comparator, and a measurement.
After projection $P$, the system ideally collapses to the desired state
\begin{equation}
|\psi\rangle=\frac{1}{C}\sum_{n=1}^{N}n^{-\beta/2}|n\rangle
\end{equation}
with ideal success possibility
\begin{equation}
    \frac{\textstyle\sum\limits_{n=1}^{N}n^{-\beta}}{\textstyle\sum\limits_{n=1}^{n_1-1}n^{-\beta}}>\frac{N}{n_1-1}\geq\frac{N}{2(N-n_0)+n_0-1}\geq \frac{N}{2N-3}>\frac{1}{2},
\end{equation}
since $n^{-\beta}$ is positive and decreasing for $\beta>0$. \par
In practice, the extended state is  prepared  to precision $\frac{\varepsilon}{6}$.  Let $|\psi_0\rangle$ and $|\widetilde{\psi_0}\rangle_e$ denote the ideal and actual extended state. Then the ideal and actual  post-selected states can be written as 
  \begin{equation}
      |\psi_0\rangle=\frac{P|\psi_0\rangle_e}{\sqrt{\langle\psi_0|_e P |\psi_0\rangle_e}}, \qquad |\widetilde{\psi_0}\rangle=\frac{P|\widetilde{\psi_0}\rangle_e}{\sqrt{\langle\widetilde{\psi_0}|_e P |\widetilde{\psi_0}\rangle_e}}.
  \end{equation}
  The precision in  $|\widetilde{\psi_0}\rangle_e$ and the contractive property of $P$ imply that
  \begin{equation}\label{eq:state diff}
      \| \sqrt{\langle\psi_0|_e P |\psi_0\rangle_e }|\psi_0\rangle-\sqrt{\langle\widetilde{\psi_0}|_e P |\widetilde{\psi_0}\rangle_e} |\widetilde{\psi_0}\rangle\|=\|P|\psi_0\rangle_e-P|\widetilde{\psi_0}\rangle_e\|\leq \||\psi_0\rangle_e-|\widetilde{\psi_0}\rangle_e\| \leq\tfrac{\varepsilon}{6}.
  \end{equation}
Furthermore, using the triangle inequality
  \begin{equation}
      |\langle\psi_0|_e P |\psi_0\rangle_e -\langle\widetilde{\psi_0}|_e P |\widetilde{\psi_0}\rangle_e|\leq |\langle\psi_0|_e P |\widetilde{\psi_0}\rangle_e -\langle\widetilde{\psi_0}|_e P |\widetilde{\psi_0}\rangle_e|+|\langle\psi_0|_e P |\psi_0\rangle_e -\langle\psi_0|_e P |\widetilde{\psi_0}\rangle_e|\leq \tfrac{\varepsilon}{3}.
  \end{equation}
  Since $\langle\psi_0|_e P |\psi_0\rangle_e\geq \tfrac{1}{2}$, it follows that
  \begin{equation}\label{eq:poss diff}
      \Bigl|\sqrt{\langle\psi_0|_e P |\psi_0\rangle_e}-\sqrt{\langle\widetilde{\psi_0}|_e P |\widetilde{\psi_0}\rangle_e}\Bigr|=\frac{|\langle\psi_0|_e P |\psi_0\rangle_e -\langle\widetilde{\psi_0}|_e P |\widetilde{\psi_0}\rangle_e|}{\sqrt{\langle\psi_0|_e P |\psi_0\rangle_e}+\sqrt{\langle\widetilde{\psi_0}|_e P |\widetilde{\psi_0}\rangle_e}}\leq \frac{\sqrt{2}\varepsilon}{3}.
  \end{equation}
  Using the triangle inequality and combining the Equations~\eqref{eq:state diff} and~\eqref{eq:poss diff},
  \begin{align}
     &\sqrt{\langle\psi_0|_e P |\psi_0\rangle_e}\||\psi_0\rangle-|\widetilde{\psi_0}\rangle\|
     =\| \sqrt{\langle\psi_0|_e P |\psi_0\rangle_e }|\psi_0\rangle-\sqrt{\langle\psi_0|_e P |\psi_0\rangle_e}|\widetilde{\psi_0}\rangle\|\notag\\
     \leq & \| \sqrt{\langle\psi_0|_e P |\psi_0\rangle_e} |\psi_0\rangle-\sqrt{\langle\widetilde{\psi_0}|_e P |\widetilde{\psi_0}\rangle_e} |\widetilde{\psi_0}\rangle\|+\| \sqrt{\langle\psi_0|_e P |\psi_0\rangle_e }|\widetilde{\psi_0}\rangle-\sqrt{\langle\widetilde{\psi_0}|_e P |\widetilde{\psi_0}\rangle_e }|\widetilde{\psi_0}\rangle\|\notag\\
     \leq&\frac{\varepsilon}{6}+\frac{\sqrt{2}\varepsilon}{3}.
  \end{align}
  Given $\bra{\psi_e} P \ket{\psi_e} \geq \frac{1}{2}$, we have the stated bound
  \begin{equation}
      \||\psi_0\rangle-|\widetilde{\psi_0}\rangle\|\leq \tfrac{1+2\sqrt{2}}{3\sqrt{2}}\varepsilon<\varepsilon,
  \end{equation}
  and success possibility satisfies 
  \begin{equation}
      \langle\widetilde{\psi_0}|_e P |\widetilde{\psi_0}\rangle_e\geq\frac{1}{2}-\frac{\varepsilon}{3}.
  \end{equation}
Thus, the ideal state $|\psi_0\rangle$ can be prepared with precision $\varepsilon$ with at least the success probability as claimed. \qed
\section{Construction of the evolution operator}
We now demonstrate the construction of the evolution operator.
\begin{theorem}[Evolution operator construction (Methods, Theorem~4)]\label{Theorem:Hsi}
    Define the time evolution operator
    \begin{equation}
        U(t)=\mathrm{e}^{-iH_0t}, \qquad H_0=\sum_{n=1}^{N}\log(n)|n\rangle\langle n|,
    \end{equation}
    where the evolution time $t$ is specified with $u$ significant digits. Then $U(t)$ can be implemented to precision $\xi$ using 
    \begin{equation}
        \mathrm{Poly}(\log(N),\log(|t|),\log(1/\xi),u)
    \end{equation}
 gates and  ancilla qubits.
\end{theorem}
\noindent\textit{Proof.} Consider an input state of the form
\begin{equation}
    \sum_{n=1}^{N} \alpha_n |n\rangle |0\rangle,
\end{equation}
where $\ket{0}$ represents the ancilla register. Applying the logarithm oracle $L$ produces
\begin{equation}
    \sum_{n=1}^{N} \alpha_n |n\rangle |\log(n)\rangle.
\end{equation}
Multiplying $\log(n)$ by $t$ yields
\begin{equation}
    \sum_{n=1}^{N} \alpha_n |n\rangle |\log(n)t\rangle.
\end{equation}
Applying controlled $R_z$ rotations results in
\begin{equation}
    \sum_{n=1}^{N} \alpha_n \mathrm{e}^{-i\log(n)t}\,|n\rangle |\log(n)t\rangle.
\end{equation}
Uncomputing the multiplication and applying inverse logarithm oracle $L^{-1}$ leaves
\begin{equation}
    \sum_{n=1}^{N} \alpha_n \mathrm{e}^{-i\log(n)t}\,|n\rangle |0\rangle \;=\; U \sum_{n=1}^{N} \alpha_n |n\rangle |0\rangle.
\end{equation}
In practice, the logarithm $\log(n)$ is approximated by $\widetilde{\log(n)}$ with error at most $\xi/|t|$. This  ensures that
\begin{equation}
   \Bigl\| \sum_{n=1}^{N} \alpha_n \bigl(\mathrm{e}^{-i\widetilde{\log(n)}\,t} - \mathrm{e}^{-i\log(n)t}\bigr)|n\rangle \Bigr\|
   \leq 2\sin\!\left(\frac{\xi}{2}\right)
   <\xi.
\end{equation}
Thus, the distance between the achieved and ideal $U$ is bounded by $\xi$. By Lemma~\ref{the:log}, the logarithm oracle and its inverse can be realized using $\log^3(N)\log^2(|t|/\xi)$ gates and $\log^2(N)\log(|t|/\xi)$ ancilla qubits.  
Multiplying $\log(n)$ by $t$ requires $(\log(|t|/\xi)+\log(N))\,u$ gate operations and $(\log(|t|/\xi)+\log(N))+u$ ancilla qubits.  
The controlled $R_z$ gates can be implemented qubit by qubit with $\mathcal{O}(\log(N)+\log(|t|/\xi))$ gates. Combining these bounds, the total resource requirement for implementing $U(t)$ is
\begin{equation}
    \mathrm{Poly}\!\left(\log(N),\,\log(|t|),\,\log(1/\xi),\,u\right)
\end{equation}
gates and ancilla qubits, completing the proof. \qed
\section{Upper bound of $\zeta'(s)$ for $0<\beta<1$}
We establish upper bounds for the Dirichlet eta function \(\eta(s)\) and its derivative \(\eta'(s)\), defined as
    \begin{equation}
    \eta(s)=\sum\limits_{n=1}^\infty (-1)^{n+1}n^{-s},\qquad \eta^{'}(s)=-\sum\limits_{n=1}^\infty (-1)^{n+1}\ln n\:n^{-s}.
\end{equation}
Then $\eta(s)$ can be expressed as:
\begin{equation}
  \eta(s)=\sum_{n=1}^{\infty}(-1)^{n+1}n^{-s}=\sum_{n=1}^{\infty}((2n-1)^{-s}-(2n)^{-s})=\sum_{n=1}^{\infty}s\int_{2n-1}^{2n}\nu^{-s-1}d\nu.  
\end{equation}
Taking the absolute value, we obtain
\begin{equation}
    |\eta(s)|\leq |s| \sum_{n=1}^{\infty}\int_{2n-1}^{2n} \nu^{-\beta-1}d\nu<|s|\int_{1}^{\infty} \nu^{-\beta-1}d\nu=\frac{|s|}{\beta}.
\end{equation}
Similarly, for the derivative \(\eta'(s)\), we have
\begin{equation}
  \eta^{'}(s)=-\sum_{n=1}^{\infty}(-1)^{n+1}(\ln n)n^{-s}=\sum_{n=1}^{\infty}(\ln{(2n)}(2n)^{-s}-\ln{(2n-1)}(2n-1)^{-s})=\sum_{n=1}^{\infty}\int_{2n-1}^{2n}(-s\nu^{-s-1}\ln{\nu}+\nu^{-s-1})d\nu.
\end{equation}
Hence,
\begin{equation}
    |\eta^{'}(s)|\leq |s|\int_{1}^{\infty}|\nu^{-s-1}|\ln{\nu} \,d\nu+\int_{1}^{\infty}|\nu^{-s-1}|\,d\nu=|s|\int_{1}^{\infty}\nu^{-\beta-1}\ln{\nu}\,d\nu+\int_{1}^{\infty}\nu^{-\beta-1}\,d\nu=\frac{|s|}{\beta^2}+\frac{1}{\beta}.
\end{equation}
Since the  zeta function is related to \(\eta(s)\) by
\begin{equation}
    \zeta(s)=\frac{1}{1-2^{1-s}}\sum_{n=1}^\infty\frac{(-1)^{n+1}}{n^s}=\frac{1}{1-2^{1-s}}\eta(s).
\end{equation}
Its derivative \(\zeta'(s)\) is:
\begin{equation}
    \zeta^{'}(s)=\frac{d}{ds}(\frac{1}{1-2^{1-s}}\sum_{n=1}^\infty \frac{(-1)^{n+1}}{n^s})
    =\frac{\eta^{'}(s)(1-2^{1-s})-2^{1-s}\ln{2}\:\eta(s)}{(1-2^{1-s})^2}.
\end{equation}
Using the previously established bounds
\[
|\eta(s)| \leq \frac{|s|}{\beta}, 
\qquad
|\eta'(s)| \leq \frac{|s|}{\beta^2}+\frac{1}{\beta},
\]
we obtain the bound for \( \zeta'(s) \):
\begin{equation}
    |\zeta^{'}(s)|\leq \frac{|s|}{|2^{1-\beta}-1|\beta}+\frac{2^{1-\beta}\ln{2}(|s|\beta+1)}{|2^{1-\beta}-1|^2\beta^2}
    =\mathcal{O}(\frac{|s|}{\beta|1-\beta|}+\frac{|s|\beta+1}{|1-\beta|^2\beta^2})
    = \mathrm{Poly}(|t|,|1-\beta|^{-1},\beta^{-1}),
\end{equation}
given that $|2^{1-\beta}-1|\geq \ln(2)|1-\beta|$ for all $0<\beta<1$.
\section{Order of $|\chi(s)|$ for $0<\beta<1$}
Recall that 
\begin{equation}
    \chi(s)=2^s\pi^{s-1}\sin\left(\frac{\pi s}{2}\right)\Gamma(1-s).
\end{equation}
Thus,
\begin{equation}
    |\chi(s)|=2^{\beta}\pi^{\beta-1}|\sin\frac{\pi s}{2}||\Gamma(1-s)|.
\end{equation}
We focus on \(s = \beta + it\) with $0<\beta<1$, 
\begin{equation}
    \sin(\tfrac{\pi s}{2} )=\sin(\tfrac{\pi \beta}{2}+\tfrac{i\pi t}{2}  )=\sin (\tfrac{\pi\beta}{2})\cosh (\tfrac{\pi t}{2})+i\cos (\tfrac{\pi\beta}{2})\sinh (\tfrac{\pi t}{2}),
\end{equation}
so we obtain
\begin{equation}
|\sin(\pi s)|=\sqrt{\sin^2 (\tfrac{\pi\beta}{2})\cosh^2 (\tfrac{\pi t}{2})+\cos^2(\tfrac{\pi\beta}{2})\sinh^2(\tfrac{\pi t}{2})}=\sqrt{\sinh^2(\tfrac{\pi t}{2})+\sin^2(\tfrac{\pi\beta}{2})}=\Theta(\mathrm{e}^{\pi|t|/2}).
\end{equation}
Using Stirling’s formula in the sector $|\arg z|\leq \pi-\varepsilon$,
\begin{equation}
    \Gamma(z)=\sqrt{2\pi}\,z^{\,z-\tfrac{1}{2}}\mathrm{e}^{-z}\bigl(1+O(|z|^{-1})\bigr). 
\end{equation}
Taking absolute values gives 
\begin{equation}
    |\Gamma(z)|=\sqrt{2\pi}\,|z^{\,z-\tfrac{1}{2}}|\mathrm{e}^{-\Re(z)}\bigl(1+O(|z|^{-1})\bigr).
\end{equation}
Note that
\begin{equation}
    |z^{z-\frac{1}{2}}|=\exp\bigl(\Re(\ln(z)(z-\tfrac{1}{2}))\bigr)=\exp\bigl(\ln(|z|)(\Re(z)-\tfrac{1}{2})-\arg(z)\Im(z)\bigr)
    =|z|^{\Re(z)-\tfrac{1}{2}}\mathrm{e}^{-\arg(z)\Im(z)}.
\end{equation}
Now we set $z=1-s=(1-\beta)-it$, then
\begin{align}
|\Gamma(1-s)|
&=\sqrt{2\pi}\,|1-s|^{1/2-\beta}\mathrm{e}^{t\arg(z)}\mathrm{e}^{-(1-\beta)}(1+O(|t|^{-1})\bigr).
\end{align}
For $t<0$, $\arg z=\arctan(\tfrac{|t|}{1-\beta})$, so 
\begin{equation}
    \exp(t\arg z)=\exp(t\:\mathrm{arccot}\tfrac{1-\beta}{|t|})=\exp\bigl(t(\tfrac{\pi}{2}-\tfrac{1-\beta}{|t|}+\mathcal{O}(|t|^{-3}))\bigr)=\exp(-\pi |t|/2)\exp((1-\beta))(1+\mathcal{O}(|t|^{-2})).
\end{equation}
For $t>0$, $\arg (z)=-\arctan(\tfrac{t}{1-\beta})$, so 
\begin{equation}
    \exp(t\arg (z))=\exp(-t\:\mathrm{arccot}\tfrac{1-\beta}{t})=\exp\bigl((-t(\tfrac{\pi}{2}-\tfrac{1-\beta}{t}+\mathcal{O}(|t|^{-3}))\bigr)=\exp(-\pi t/2)\exp((1-\beta))(1+\mathcal{O}(t^{-2})).
\end{equation}
In both cases,
\begin{equation}
    \exp(t\arg z)=\exp(-\pi |t|/2)\exp((1-\beta))(1+\mathcal{O}(|t|^{-2})).
\end{equation}
Therefore,
\begin{equation}
    |\Gamma(1-s)|=\sqrt{2\pi}\,|1-s|^{1/2-\beta}\exp(-\pi |t|/2)(1+O(|t|^{-1})\bigr).
\end{equation}
Combining terms,
\begin{equation}
    |\chi(s)|=\Theta(1)2^{\beta}\pi^{\beta-1}\exp(\pi|t|/2)\sqrt{2\pi}|1-s|^{1/2-\beta}\exp(-\pi |t|/2)(1+O(|t|^{-1})),
\end{equation}
The exponential terms cancel, leaving
\begin{equation}
    |\chi(s)|=\Theta(|1-s|^{1/2-\beta})=\Theta(|t|^{1/2-\beta}).
\end{equation}

\end{document}